\documentclass{aa}  

\usepackage{graphicx}
\usepackage{natbib}
\usepackage{txfonts}
\usepackage[fleqn]{amsmath}
\newcommand{\beq}{\begin{equation}}
\newcommand{\eeq}{\end{equation}}
\newcommand{\bea}{\begin{eqnarray}}
\newcommand{\eea}{\end{eqnarray}}
\newcommand{\rem}[1]{ }
\newcommand{\F}{\mathcal F}

\begin{document}

\title{Exponentially growing bubbles around early supermassive black holes}

   \author{
   R.~Gilli\inst{1},
   F.~Calura\inst{1},
   A.~D'Ercole\inst{1}
          \and
          C.~Norman\inst{2,3}
          }

   \institute{INAF -- Osservatorio Astronomico di Bologna, via Gobetti 93/3, 40129 Bologna, Italy\\
              \email{roberto.gilli@oabo.inaf.it}
         \and
        Department of Physics and Astronomy, Johns Hopkins University, Baltimore, MD 21218, USA
        \and
        Space Telescope Science Institute, 3700 San Martin Drive, Baltimore, MD 21218, USA
             }

   \date{Received; accepted}

% \abstract{}{}{}{}{} 
% 5 {} token are mandatory
 
  \abstract
{
%Supermassive black holes (SMBHs) at galaxy centers grow during active phases in which they both efficiently accrete matter and release
%radiation and matter into the surrounding environment. As an example, the growth of early seeds to the billion solar mass BHs powering QSOs
%at $z~6$ must have been exponential and uninterrupted, lasting a few hundreds Myr at least. If a fixed fraction of the QSO bolometric
%output were converted into a gas outflow, this may also have grown exponentially.

We address the as yet unexplored issue of outflows induced by exponentially growing power sources, focusing on early supermassive black holes (BHs). 
We assumed that these objects grow to $10^9\;M_{\odot}$ by z=6 by Eddington-limited accretion and convert 5\% of their bolometric output into a wind.
We first considered the case of energy-driven and momentum-driven outflows expanding in a region where the gas and total mass densities are uniform and equal to the average
values in the Universe at $z>6$. We derived analytic solutions for the evolution of the outflow:   for an exponentially growing power with e-folding time $t_{Sal}$, 
we find that the late time expansion of the outflow radius is also exponential, with e-folding time of $5t_{Sal}$ and $4t_{Sal}$ in the energy-driven and momentum-driven limit, respectively.

We then considered energy-driven outflows produced by quasi-stellar objects (QSOs) at the centre of early dark matter halos of different masses and powered by BHs growing from
different seeds. We followed the evolution of the source power and of the gas and dark matter density profiles in the halos from the
beginning of the accretion until $z=6$. The final bubble radius and velocity do not depend on the seed BH mass, but are instead smaller for larger halo masses. At z=6, bubble
radii in the range 50-180 kpc and velocities in the range 400-1000 km~s$^{-1}$ are expected for QSOs hosted by halos in the mass range $3\times10^{11}-10^{13}\;M_{\odot}$.
These radius and velocity scales compare well with those measured for the outflowing gas in the z=6.4 QSO SDSS~J1148+5251.

%We assumed that the gas in the halo is at the virial temperature. For large enough halos, where the gas temperature and sound speed are higher,  the expansion of the 
%bubble may become subsonic in a given time interval. We find that for halos with $M_h\leq3\times10^{11}\;M_{\odot}$ at z=6 the outflow is always supersonic. The fraction of time spent at subsonic velocities increases for larger masses, until it is always subsonic for $M_h\geq5\times10^{12}\;M_{\odot}$ at z=6. We also explored the effects of a 
%assuming lower ambient gas temperature, down to $T_{vir}/4$. 
%We found that the transitional halo masses between fully subsonic and fully supersonic outflows do not strongly depend on the assumed gas temperature and corresponding density profile. 
%For lower temperatures / steeper profiles the bubble radii and velocities can reach values up to 1 Mpc and a few $\times 1000$ km~s^{-1}, respectively.

By the time the QSO is observed, we found that the total thermal energy injected within the bubble in the case of an energy-driven outflow is $E_{th}\sim5 \times 10^{60}$ erg.
This is in excellent agreement with the value of $E_{th}=(6.2\pm 1.7) \times 10^{60}$ erg measured  through the detection of the thermal 
Sunyaev-Zeldovich effect around a large population of luminous QSOs at lower redshifts. This suggests that QSO outflows are closer to the energy-driven limit than to the momentum-driven limit.

We investigated the stability of the expanding gas shell in the case of an energy-driven supersonic outflow propagating within a dark matter halo with $M_h=3\times10^{11}\;M_{\odot}$
at z=6. We found that the shell is Rayleigh-Taylor unstable already at early times and, by means of a simple model, we investigated the fate of the fragments detaching from the shell. We found that these fragments should rapidly evaporate because of the extremely high temperature of the hot gas bubble if this does not cool. Since the only effective cooling mechanism for such a gas is inverse Compton by the cosmic microwave background (CMB) photons (IC-CMB), which is important only at $z\geq 6$, we speculate that such shell fragments may be observed only around high-z QSOs, where IC-CMB cooling of the bubble gas can prevent their evaporation.

%We finally propose that those shell fragments that have already fallen back towards the center of the dark matter halo by the time we
%observe the QSO may accumulate and constitute a reservoir
%of relatively cold gas ($T\sim 10^4$ K) around QSOs on scales of up to a few tens kpc. This mechanism could explain the ubiquitous presence of such a gas observed by MUSE
%around $z\sim3.5$ QSOs.
}

   \keywords{black hole physics -- quasars: supermassive black holes -- shock waves -- galaxies: high-redshift}

   \maketitle

\section{Introduction}

Active galactic nuclei (AGN) may drive powerful gas outflows out to distances much greater than kpc. Observations have shown that the outflowing gas may be in a broad 
range of ionization states and may cover large solid angles. For instance, the detection of blueshifted metal absorption lines in the X-ray spectra of nearby AGN revealed the 
presence of ultra-fast outflows (UFOs; see e.g. \citealt{tombesi12, tombesi13} and references therein) produced within a few milli-pc of the central black hole (BH). 
Ultra-fast outflows are made of plasma moving at 0.1c on average, and up to $\sim0.3$c \citep{tombesi15, nardini15}. Although it is difficult to constrain the opening angle of UFOs, in the powerful local quasi-stellar object (QSO) PDS456 the remarkable P Cygni profile of the FeXXVI $K\alpha$ line indicates the presence of a subrelativistic wind expanding almost spherically. On much larger scales ($\sim$kpc), outflows of less ionized gas have been traced through observations of asymmetric emission lines in the optical regime, such as the [OIII]5007\AA\ line \citep{brusa15, harrison16, shen16, zakamska16}. Outflows of neutral atomic and molecular gas have also been  observed. 
The neutral atomic component has been probed by observations of the Na I D absorption doublet \citep{krug10} or by the [CII]158$\mu$m and HI 21cm fine structure emission lines \citep{maiolino12, morganti16}. The molecular component is mostly probed by observations of CO transitions \citep{feruglio10, cicone14}. Velocities higher than 1000 km~s$^{-1}$ are commonly observed in all of these gas phases. The molecular component generally dominates the total mass budget of the large-scale outflowing gas.
The energy required to drive  large-scale outflows in AGN is so large that in most cases it can only be provided by  nuclear activity rather than by  star formation in the host \citep{cicone14}. In at least two QSOs, namely Mrk231 \citep{feruglio15} and IRAS~F11119 \citep{tombesi15},  a UFO and a molecular outflow have both been observed. The
comparison between the energy of the inner UFOs and that of the large-scale molecular outflows suggest that AGN outflows are closer to the  energy-driven 
limit than to the momentum-driven limit (see Section 2); that is, the UFOs are able to inflate a bubble of hot gas that does not cool efficiently and whose thermal pressure
can push layers of colder gas to large distances. Very recently, further support for the existence of hot gas bubbles around AGN came from the 
detection of the thermal Sunyaev-Zeldovich effect on the stacked far-infrared spectra of large populations of $z\sim 2$ QSOs \citep{ruan15, crichton16}

AGN-driven outflows are obviously best studied in nearby objects, but they are observed in AGN at all redshifts \citep{brusa15, zakamska16}, even beyond redshift 6 \citep{maiolino12}, i.e. in the most distant QSOs known. 
A remarkable example is given by the famous QSO SDSSJ1148+5251 \citep{fan03}, one of the first $z>6$ QSOs discovered by the SDSS survey, where observations of the 
[CII]158$\mu$m line with the PdBI interferometer have revealed the presence of large-scale gas outflows extending out to 30 kpc from the QSO and moving with velocities of 
$>1000$ km~s$^{-1}$ \citep{cicone15}. This large-scale outflow extends well beyond the characteristic size of the host galaxies of $z\sim6$ QSOs ($\sim1-3$ kpc \citealt{wang13,venemans17}), and then expands on scales comparable with that of the hosting dark matter halo, possibly affecting the QSO local environment.

The theory of QSO driven outflows has been presented in many works, either using an analytic/numerical approach \citep{king10, king11, zubovas12, fg12},  detailed hydrodynamical simulations \citep{nayakshin12, costa14},
or a combined approach linking simulations at different scales through analytic recipes \citep{gaspari17}.
A common assumption in most of these works is that the source is constant in power and that the outflow expands within a dark matter halo potential and gas density profile
which are constant. In general, despite the vast literature on outflows produced by astrophysical sources (supernovae- , star- , AGN-driven winds), only energy sources
that are impulsive, constant, or evolve as power laws have been considered.
In this work we explore the issue of outflows produced by exponentially growing power sources, such as should be the case in early QSOs. In fact, to explain
the billion solar mass black holes observed in QSOs at $z>6$ \citep{willott10bis,derosa14}, i.e. when the Universe was less than 1 Gyr old, black holes must have grown exponentially by
accreting continuously at the Eddington limit (but see e.g. \citealt{madau14,volonteri15,pezzulli16} for intermittent, super-critical growth of early black holes). If a fixed fraction of the QSO power is transferred to an inner UFO, then the power of this wind must also grow exponentially. Although assuming continuous and efficient BH accretion for several e-folding times is one of the few ways of coping with the still unsolved issue of
the formation of the first supermassive black holes (SMBHs), we note here that 
this is inconsistent with simulations of BH growth at high Eddington ratios. Pre-heating instabilities that halt BH fueling already arise  at Eddington ratios of $\sim 0.01$ in the case of spherical accretion \citep{cos78}.
Hydrodynamic simulations in 2D have shown that this also occurs when removing the constraint of spherical accretion \citep{novak11, ciotti12}. At high Eddington ratios an additional instability effect is expected
to take place, due to a recurrent thickening of the accretion disc and the consequent increase of its covering factor. Because of these arguments, uninterrupted BH accretion is therefore unlikely,  yet it can be regarded as a useful approximation of the long-term behaviour of a more erratic growth.
Finally, we note that at high redshifts the mass growth of the hosting dark matter halo and the change in the gas profile within the halo are not negligible during the accretion time of the QSO, so we followed 
them self-consistently during the outflow expansion, in contrast  to most previous works.

A concordance cosmology with $H_0=70$ km~s$^{-1}$~Mpc$^{-1}$ , $\Omega_m=0.3$ , $\Omega_\Lambda=0.7$, in agreement within the errors with the Planck 2015 results \citep{planck16}, 
is used throughout this paper.

\section{Problem set-up and definitions}

We consider the case of an accreting supermassive black hole radiating at a given fraction $\lambda\equiv L_{bol}/L_E$ of its  Eddington luminosity 
$L_E=4\pi G m_p c M/\sigma_T =M c^2/t_E=1.26 \times 10^{38} M$ erg~s$^{-1}$, where $M$ is the black hole mass in units of $M_{\odot}$ and
$t_E=\sigma_T c/(4\pi G m_p)=0.45$ Gyr is the Eddington time.

If a fraction $\epsilon$ of the mass $m_{acc}$ falling onto the black hole is converted into radiation, then $L_{bol}=\epsilon \dot m_{acc} c^2 $, with 
$\dot m_{acc} = \lambda L_E/(\epsilon c^2)=\lambda M/(\epsilon t_E)$. The black hole growth rate is then
$\dot M = (1-\epsilon) \dot m_{acc} = (1-\epsilon)\lambda M / (\epsilon t_E)$ ,
from which it follows that

\beq
M(t)=M_0 e^{t/t_{Sal}} ,
\eeq

where 

\beq
t_{Sal}=\frac{\epsilon}{1-\epsilon}\frac{t_E}{\lambda} = 50 \; Myr \; \left(\frac{9\epsilon}{1-\epsilon}\right)\lambda^{-1} 
\eeq is the Salpeter (e-folding) time, and $M_0$ is the mass of the black hole ``seed''.
Under these assumptions, the bolometric QSO luminosity grows as

\beq
L_{bol}(t) = \lambda L_E = \lambda \frac{M(t) c^2}{t_E} =  L_{bol,0} e^{t/t_{Sal}} ,
\eeq where $L_{bol,0} = \lambda c^2 M_0 / t_E$. It is assumed that a constant fraction $f_{w}$ of the AGN bolometric luminosity powers a supersonic wind pushing on the surrounding ambient medium.
This fraction is typically considered to be $f_{w}$=0.05 \citep{scanna04,lapi05,germain09}.
The outflow is also assumed to be almost spherical (i.e. a bubble) with covering factor $\Omega \approx 4\pi$.
We then investigate the evolution of astrophysical bubbles produced by an exponentially growing input energy source,

\beq
L_{w}(t) =  \frac{1}{2} \dot m_{w} v_{w}^2 = L_{w,0} \;e^{t/t_{Sal}}= f_{w} L_{bol,0}\;e^{t/t_{Sal}}\; ,
\label{input}
\eeq where $\dot m_{w}$ and $v_{w}$ are the wind mass rate and velocity, respectively.

Observations of ultra-fast  outflows in the X-ray spectra of QSOs (e.g. \citealt{tombesi13}) suggest that the gas outflow
rate is of the order of the gas accretion rate onto the BH, so we  assume $\dot m_{w}\approx \dot m_{acc}$. We further assume that the QSO wind is launched with constant velocity.\footnote{Under the above assumptions, Eq.~\ref{input} implies that $2f_w\epsilon(c/v_w)^2=1$, i.e. $v_w=0.1c$ for $f_w=0.05$ and $\epsilon=0.1$ as assumed here, 
in agreement with the velocities measured for UFOs.}
Therefore, the outflow rate grows as

\beq
\dot m_{w}(t) = \dot m_{w,0}e^{t/t_{Sal}} ,
\label{mdot}
\eeq where

\beq
\dot m_{w,0}=\frac{\lambda M_0}{\epsilon t_E} = 2.2 \times 10^{-4} \; M_{\odot}\; yr^{-1} \; \lambda \left(\frac{0.1}{\epsilon}\right)\left( \frac{M_0}{10^4 M_{\odot}}\right).
\eeq

In turn, the ejected wind mass grows as

\beq
m_{w}(t)=\int_0^t \dot m_{w}dt^{\prime} = \dot m_{w,0} t_{Sal} (e^{t/t_{Sal}}-1) .
\label{ewm}
\eeq

\begin{figure}[t]
\includegraphics[angle=0, width=9cm]{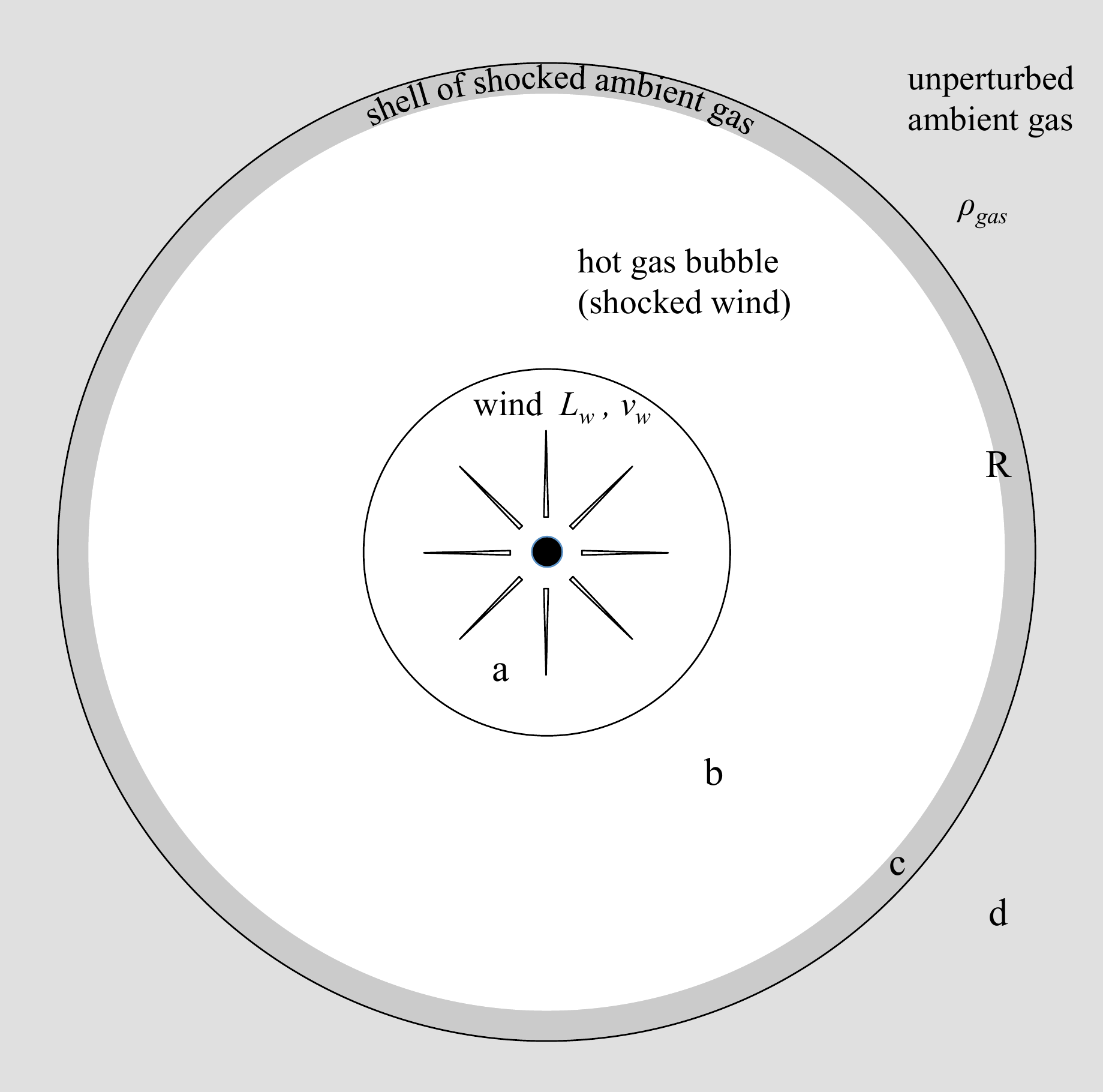}
\caption{Classic structure of a spherical outflow produced by a fast, supersonic wind colliding into an ambient medium (see also  \citealt{w77,dw92,costa14}). The flow is divided into four main regions labelled  $a$, $b$, $c$, $d$. Region $a$ is occupied by the QSO wind launched at a velocity
$v_w$ and carrying a kinetic power $L_w$. A reverse shock between regions $a$ and $b$ shocks and heats the QSO wind. It is the cooling of the shocked wind
in region $b$ (the bubble) that determines whether the outflow is energy-driven (no cooling) or momentum-driven (instantaneous cooling). 
Region $c$ is the thin shell of shocked ambient gas in pressure equilibrium with the gas in region $b$ through a contact discontinuity. Region $d$ shows the still unperturbed ambient medium with density $\rho_{gas}$ where the outflow expands.
} 
\label{ofs}
\end{figure}

\section {Energy-driven vs. momentum-driven outflows and outflow structure}

The classic theory of astrophysical bubbles driven by fast winds (see \citealt{w77} and \citealt{omk88} for reviews) shows that
the structure of the outflow is characterized by a forward and a reverse shock, and is divided into four main regions (see Fig. \ref{ofs}).
The forward shock propagates in the ambient medium (region $d$ in Fig. \ref{ofs}), enhancing its density in a 
shell (region $c$) that progressively sweeps up more gas. The outflowing wind  (region $a$) instead encounters the innermost reverse shock, and is therefore heated.
The region $b$ of hot shocked wind is usually termed a ``bubble''. The cooling properties of the shocked wind define whether an outflow is energy-
or momentum-driven. In the limit in which this gas does not cool,  all the energy originally provided by the wind is retained within
the system, and we refer to the outflow as ``energy-driven'' \footnote{Here we  neglect the cooling of the gas within the shell. This would,  in any case, 
 affect the estimates of the bubble radius and velocity by less than 15\% \citep{w77}.}. It is the thermal pressure of the hot gas within the bubble that accelerates
the shell of gas into the ambient medium. On the contrary, if all thermal energy within the bubble is radiated away instantaneously,
energy is not conserved, whereas all the momentum carried by the wind is transferred directly to the shell of shocked ambient gas.
The outflow is therefore referred to as ``momentum-driven'', which effectively corresponds to maximal cooling. 
Realistic outflows obviously fall between these two limits. It is, however, instructive to treat them separately.

We follow an approach similar to that of \citet{w77}, who provided the basics for treating the structure and evolution of bubbles
inflated by fast stellar winds within a medium of constant density $\rho_{gas}$. Their computations can be generalized to astrophysical bubbles
powered by other energy sources. 
In the energy-driven case, the following equations can be combined to obtain the full equation of motion of the expanding shell:

\beq
E_{th} = \frac{3}{2}P\frac{4\pi}{3}R^3 = 2\pi P R^3 \; ,
\label{ep}
\eeq

\beq
\frac{d}{dt}[M(R)\dot R] = 4\pi R^2 P \; ,
\label{cmom}
\eeq

\beq
\dot E_{th} = L_w - 4\pi R^2 P \dot R \; .
\label{cene}
\eeq

Equation~\ref{ep} shows the relation between thermal energy $E_{th}$ and pressure $P$ for a monoatomic gas in a volume $V=(4\pi/3)R^3$.

Eq.~\ref{cmom} describes the evolution of the momentum of the shell of swept-up gas (the left-hand side is the rate of change of momentum of the shell 
and the right-hand side represents the force acting on it). 

Eq.~\ref{cene} is the energy equation of the system: the rate at which the thermal energy of the hot gas in the bubble (left-hand side) changes is equal
to the kinetic luminosity of the wind minus the rate at which the hot gas does work on the surrounding medium. In the momentum-driven case, $E_{th}=0$
and the pressure $P$ in Eq.~\ref{cmom} is given by the wind ram pressure

\beq
P = L_w/(2\pi v_w R^2) ,
\label{ram}
\eeq where $L_w$ and $v_w$ are the wind kinetic power and velocity, respectively.

\citet{w77} considered the case of a constant source of energy. Energy inputs evolving in time as power laws were considered by e.g. Koo \& McKee (1992).
Here we  consider sources with exponentially growing luminosities.

In our equations we have assumed that the effects of the pressure of the ambient gas and of gravity are both negligible.
We  explore the effects of relaxing these assumptions in the next sections. 

\section {QSO in a uniform density field}

We first assume that the BH forms and grows, and hence the outflow propagates, within a region with uniform density equal to the mean matter density of the Universe.
We recall that the cosmic evolution of the baryon density is given by

\beq
\rho_b(z) = \rho_{b,0}(1+z)^3 ,
\label{rho}
\eeq

where $\rho_{b,0}=3 H_0^2 \Omega_{b,0} /(8\pi G)$ is the mean baryon density at $z=0$. For $H_0=70$ km~s$^{-1}$~Mpc$^{-1}$ and $\Omega_{b,0}=0.045$, $\rho_{b,0}=4\times 10^{-31}$ g~cm$^{-3}$.
At $z=6$ the mean baryon density is therefore $\rho_{gas}\equiv\rho_b(6)\sim10^{-28}$ g~cm$^{-3}$.
For an ambient gas with uniform density $\rho_{gas}$, the mass of the shell of swept-up gas $M_s(R)=\int_0^R \rho_{gas}(R)dV$ is equal to $(4\pi/3) R^3 \rho_{gas}$.

\subsection{Energy-driven case}

Following \citet{dw92}, we combine Eqs. \ref{ep}, \ref{cmom}, and \ref{cene} to
obtain the following equation of motion of the shell:

\beq
 15R^2\dot R^3 + 12R^3\dot R\ddot R + R^4 \dddot R = \frac{3}{2\pi}\frac{L_w(t)}{\rho_{gas}}.
\label{dw92}
\eeq

This is a non-linear differential equation that admits more than one solution, and where not all of the solutions can be obtained by linear combination of the others.
For an exponential input energy source described by Eq. \ref{input}, the simplest (but unphysical, see below) solution to Eq.\ref{dw92} is analytic and takes the form

\beq
R(t) =  R_0 \; e^{t/(5t_{Sal})} , 
\label{exp}
\eeq with

\begin{align}
\label{r0}
R_0 = \left(\frac{375}{56\pi}\right)^{\frac{1}{5}}t_{Sal}^{\frac{3}{5}}\left(\frac{L_{w,0}}{\rho_{gas}}\right)^{\frac{1}{5}} \\ \notag & \hspace{-3.5cm} = 28.5 \;kpc \; \left(\frac{9\epsilon}{1-\epsilon}\right)^\frac{3}{5}\lambda^{-\frac{2}{5}}\left(\frac{f_w}{0.05}\right)^\frac{1}{5}\left(\frac{M_0}{10^4 M_{\odot}}\right)^\frac{1}{5}\left(\frac{\rho_{gas}}{10^{-28} g\;cm^{-3}}\right)^{-\frac{1}{5}} .
\end{align}

\begin{figure}[h]
\includegraphics[angle=0, width=9cm]{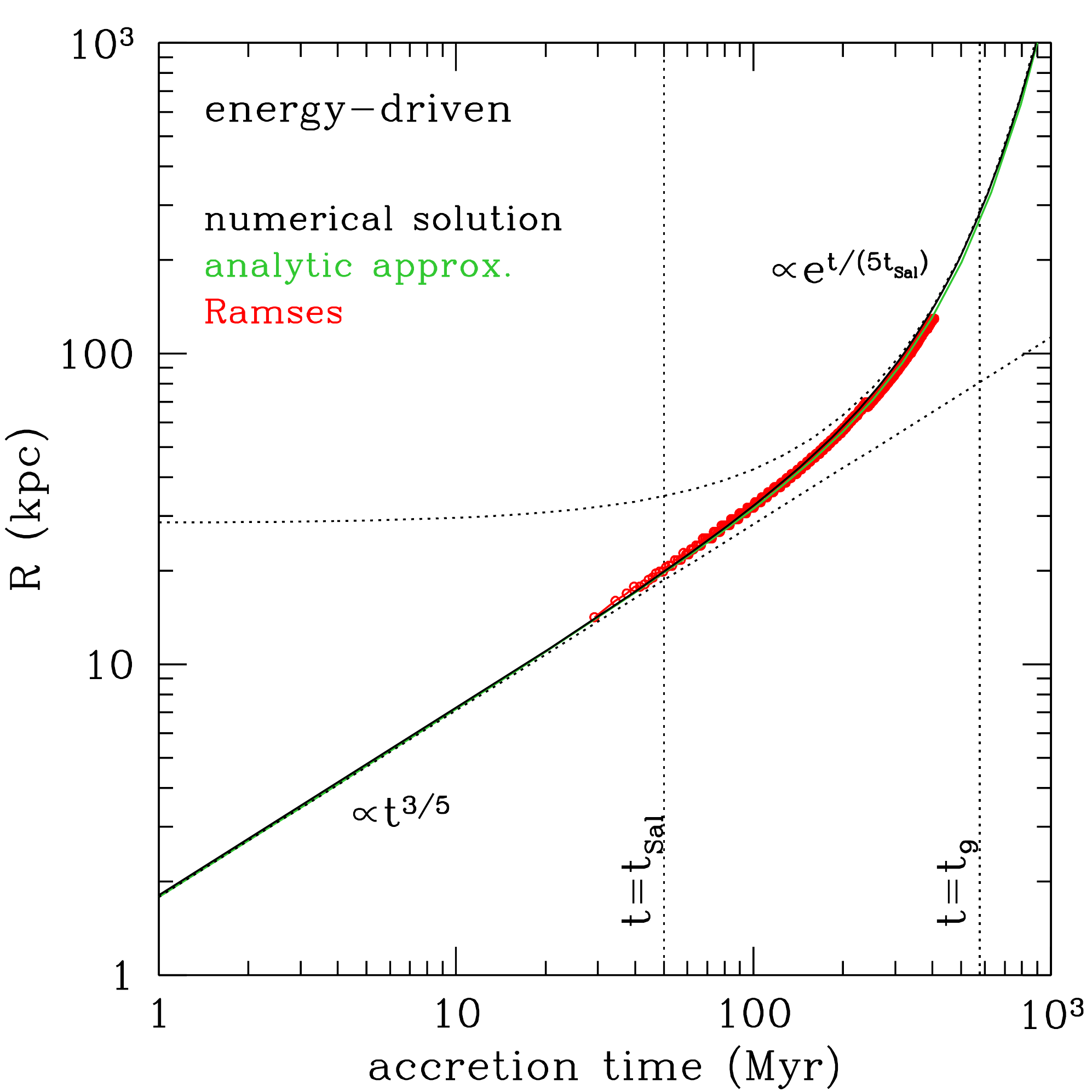}
\caption{Time evolution of the radius of a hot gas bubble inflated by a wind from an early SMBH accreting at its Eddington limit (energy-driven case). The following parameters are assumed:
  $f_w=0.05$, $\lambda=1$, $M_0=10^4\;M_{\odot}$ (hence $L_{w,0}=6.3\times10^{40}$ erg~s$^{-1}$),  $\epsilon=0.1$, 
  $\rho_{gas}=10^{-28}$ g~cm$^{-3}$. The black solid curve shows the numerical solution obtained
  for the equation of motion (Eq.~\ref{dw92}) assuming an exponentially growing wind with $L_{w}(t)=L_{w,0}e^{t/t_{Sal}}$ and assuming $R,v\rightarrow 0,0$  for $t\rightarrow 0$.
  The analytic approximation to the numerical solution (Eq.~\ref{anal}), is shown by the solid green line.
  The purely exponential (but unphysical) solution given by Eq.~\ref{exp} is shown by the black dotted curve. The black dotted  line shows the solution obtained for a constant 
  energy source $L_{w}(t)=L_{w,0}$ (Eq.~\ref{ww}). The vertical dotted lines mark the Salpeter time $t_{Sal}$ and the time $t_9$ needed to grow the SMBH to $10^9\;M_{\odot}$ as labelled.
  The bubble radius scales as $t^{3/5}$ for $t<<t_{Sal}$ and $e^{t/(5t_{Sal})}$ for $t>>t_{Sal}$.
  The red points show the results of a low-resolution simulation run with the hydrodynamical code RAMSES assuming the same input parameters. The results of the
  simulation are in excellent agreement with the numerical solution to Eq.~\ref{dw92}.
} 
\label{rvm}
\end{figure}

In Fig.~\ref{rvm} we show the time evolution of the bubble radius $R(t)$ up to the accretion time $t_9$, defined as the time at which the black hole has grown to
$10^9\;M_{\odot}$, i.e. $M(t_9)\equiv 10^9\;M_{\odot}$. This is the typical mass observed for bright SDSS QSOs at $z\sim6$. For this computation we assumed\\ \\
$f_w=0.05$ , \\
$\epsilon=0.1$ , \\
$\lambda=1$ , \\
$M_0=10^4\;M_{\odot}$ . \\
The adopted seed mass $M_0$ is  an intermediate value within the broad range predicted by different models. This range spans the whole interval from
$\sim 10^2\;M_{\odot}$, as expected for the remnants of POPIII stars \citep{mr01}, to $\sim10^4\;M_{\odot}$, as expected from the collapse of stellar clusters \citep{devecchi09}, 
and even up to $10^6\;M_{\odot}$, as postulated for large seeds forming from the direct collapse of large gas clouds under specific environmental conditions (\citealt{agarwal14}; 
see also \citealt{vb12} for a review and \citealt{valiante16} for a model of seed formation at different mass scales). With the above assumptions the initial power of the wind 
is $L_{w,0}=6.3\times10^{40}$ erg~s$^{-1}$, and $t_9\sim 580$ Myr.

\begin{figure}[h]
\includegraphics[angle=0, width=9cm]{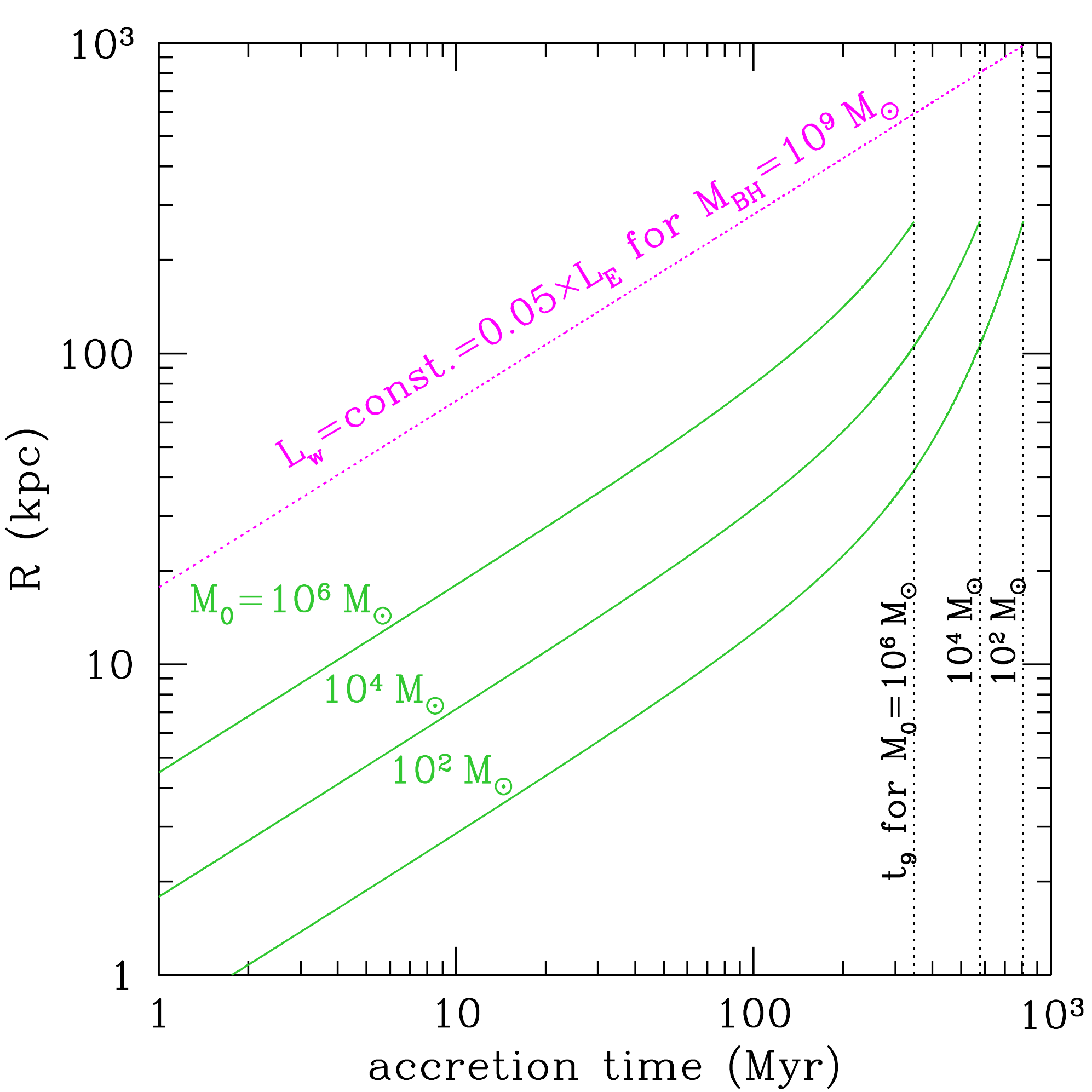}
\caption{Effects of varying the initial seed BH mass $M_0$ on the evolution of the bubble radius following Eqs.~\ref{r0} and \ref{anal} (energy-driven case).
  All other relevant parameters are the same as assumed in Fig.~\ref{rvm}. The three green solid curves are computed for $M_0=10^2\; , 10^4\; , 10^6\;M_{\odot}$ as labelled. The BH reaches
  a final mass of $10^9\;M_{\odot}$ at different times $t_9$ (shown by the dotted lines), but the bubble radius at that time, $R(t_9)$, is the same whatever is the initial seed mass.
  For reference, the magenta line shows the time evolution of a bubble inflated by a constant wind power of $L_w = f_w L_E=6.3\times10^{45}$ erg~s$^{-1}$,
  corresponding to a $10^9 M_{\odot}$ BH radiating at its Eddington limit and with $f_w=0.05$. Because of the scaling between BH mass and the normalization of the shell radius ($R_0\propto M_0^{1/5}$; Eq.~\ref{r0}), this curve has a $(10^9/10^4)^{1/5}$ = 10 times higher normalization than the black dotted line shown in Fig.~\ref{rvm}.
} 
\label{radii}
\end{figure}

\begin{figure}[h]
\includegraphics[angle=0, width=9cm]{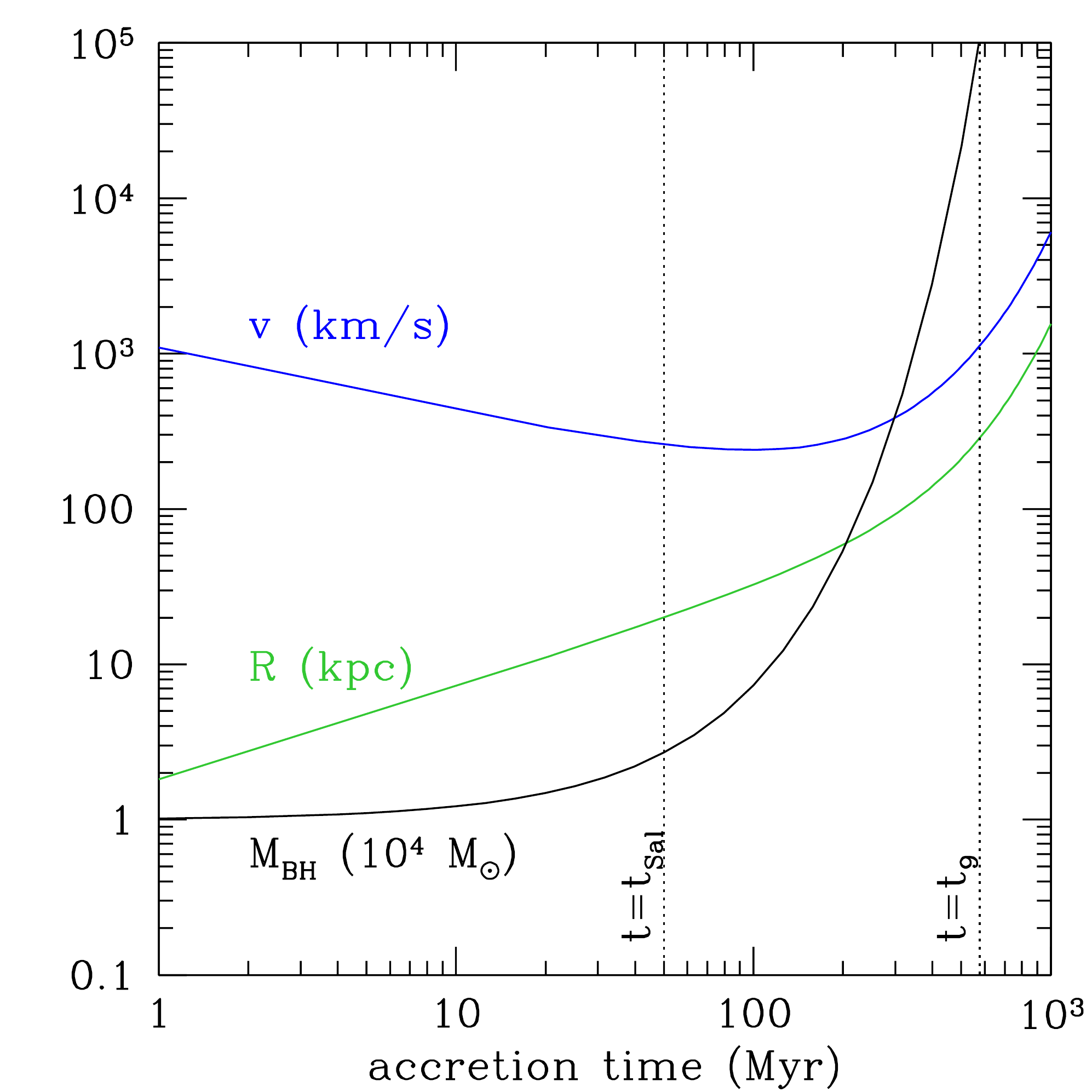}
\caption{Time evolution of the bubble radius (green curve) and velocity (blue curve) for an exponentially growing luminosity source
  powered by an exponentially growing black hole (black curve) for the energy-driven case. The source parameters and vertical dotted lines are as in Fig.~\ref{rvm}. 
  The velocity of the bubble reaches a minimum around a few Salpeter times,  after which
  the shell starts accelerating. Rayleigh-Taylor instabilities arise at this stage (see Section 5).
} 
\label{allinone}
\end{figure}

The simple solution given in Eq.~\ref{exp} would imply that at $t=0$ the radius of the bubble is $R=28.5$ kpc, i.e. the bubble ``jump starts'' on scales that
are far larger than the sphere of influence of the seed BH and even larger than its putative host galaxy. Since $R_0$ scales weakly with the seed mass
($R_0\propto M_0^{0.2}$), assuming a stellar-sized seed with $M_0=10^2\;M_{\odot}$ would only reduce $R_0$ by a factor of 2.5.
The pure exponential solution therefore appears  unphysical.

We therefore solved  Eq.~\ref{dw92} numerically by imposing $R\rightarrow 0$ (and $v\equiv \dot R \rightarrow 0$) for $t\rightarrow 0$ and obtained 
a simple analytic approximation to the numerical solution by manipulating Eq.~\ref{exp} and subtracting its second-order Taylor expansion from the exponential term

\beq
R(t)\simeq\left(\frac{8}{11}\right)^\frac{1}{5}R_0\left[e^{t/t_{Sal}}-1-\frac{t}{t_{Sal}}-\frac{1}{2}\left(\frac{t}{t_{sal}}\right)^2\right]^\frac{1}{5} .
\label{anal}
\eeq

As shown in Fig.~\ref{rvm}, Eq.~\ref{anal} represents an excellent approximation (within 5\%) of the bubble radius over the entire accretion history of the BH.
At late times ($t>>t_{Sal}$) this solution approaches the simple analytic solution given by Eq.~\ref{exp}.
At early times ($t<<t_{Sal}$), the numerical solution approaches the analytic solution that is valid for a
constant source of power $L_w(t)=L_{w,0}$ (as assumed by \citealt{w77} or \citealt{dw92}):

\beq
R(t)\simeq\left(\frac{4}{33}\right)^\frac{1}{5}R_0\left(\frac{t}{t_{Sal}}\right)^\frac{3}{5} \sim 0.65R_0\left(\frac{t}{t_{Sal}}\right)^\frac{3}{5} .
\label{ww}
\eeq

As a further check we ran a test low-resolution hydrodynamical simulation in which
a source embedded in a uniform density field injects energy and matter at an exponential rate.
The simulation was performed using a customized version of the RAMSES code \citep{teyssier02} as described by \citet{calura15}.
The initial conditions are represented by a uniform and homogeneous distribution of
gas characterized by a density of $10^{-28}$~g~cm$^{-3}$ and temperature $10^{4}$~K (see Section 4.4). \\
The computational box has a volume of $L_{box}^3 = (650$~kpc$)^3$ and is characterized by a
maximum resolution of $\sim 130$~pc.
At $t=0$, a source located in the origin of the computational box
starts injecting energy at a rate given by Eq.~\ref{input}, with $L_{w,0} = 6.3~\times~10^{40}$ erg~s$^{-1}$ and
matter at a rate given by Eq.~\ref{mdot}, with $\dot{m}_{w,0} \sim 2.2~\times~10^{-4}\;M_{\odot}$~yr$^{-1}$.
In order to avoid the occurrence of diamond-shaped shock fronts which can sometimes be present in similar conditions in cartesian
grid-based simulations \citep{tasker08}, our source is non-point-like but
distributed over a spherical volume $V=\frac{4}{3}\pi\Delta R^3$, with $\Delta R=200$~pc.
The total thermal energy and mass injected by the source in the surrounding environment per unit volume and in the time step
$\Delta t$ are $\Delta \epsilon= \dot{\epsilon} \Delta\,t$
and $\Delta \rho=\dot{\rho} \Delta\,t$, respectively, where $\dot{\rho} =\frac{\dot{m}_{w}}{V}$ and $\dot{\epsilon}=\frac{L_w}{V}$.\\
At each time step, a number of cells with the highest refinement level are created at the position of the source. 
In the remainder of the computational domain, the refinement strategy is both geometry- and discontinuity-based.
Outflow boundary conditions are used. The simulation is adiabatic, i.e. the effects of radiative cooling are assumed to be negligible.
 The effects of gravity are also neglected.
As shown in Fig.~\ref{rvm}, the expansion of the shell from the simulation is in excellent agreement with the numerical solution to Eq.~\ref{dw92}. 

At late times, Eq.~\ref{anal} is well approximated by Eq.~\ref{exp} and the growth of the bubble radius is purely exponential.
Some interesting properties can be inferred by looking at Eq.~\ref{exp}. For instance, the e-folding time of the bubble radius is five times
the Salpeter time. Also, it can be shown that the bubble radius (and velocity) at a given black hole mass (e.g. $10^9\;M_{\odot}$) does not depend on the initial
seed mass: for smaller seeds, it will simply take longer to accrete that mass and blow the bubble to that radius. This is visible in Fig.~\ref{radii}, where the
effects of varying the seed BH mass from $M_0=10^2 M_{\odot}$ to $10^6M_{\odot}$ are shown.
The bubble radius when the BH has grown to $10^9M_{\odot}$ is $R(t_9)=267$ kpc whatever the initial seed mass is.
This value is three times smaller than the radius reached in the same time span by a bubble inflated by a constant wind power of $L_w=f_w L_{bol}=f_w L_E=6.3\times10^{45}$ erg~s$^{-1}$,
corresponding to a $10^9 M_{\odot}$ BH radiating at its Eddington limit and with $f_w=0.05$. This is shown by the magenta curve in Fig.~\ref{radii}: as expected, the normalization
of this curve is a factor of  $(10^9/10^4)^{1/5}$ = 10 higher than the black dotted line in Fig.~\ref{rvm}, which was computed for a $10^4 M_{\odot}$ BH 
(again radiating at its Eddington limit and with $f_w=0.05$).
As a matter of fact, previous works on QSO outflows do not consider the increase in the QSO luminosity and wind
power as the BH grows, but rather assume a constant release of energy. 

By solving numerically Eq.~\ref{dw92} we also derived a solution for the time evolution of the bubble velocity $v\equiv \dot R$, which is compared in Fig.~\ref{allinone} with the
time evolution of the bubble radius and BH growth. The numerical solution for the bubble velocity can be approximated (within $\sim$6\%) by the derivative of the analytic approximation
to the bubble radius given in Eq.~\ref{anal}:

\beq
v(t) \simeq \frac{8}{55}\frac{R_0}{t_{Sal}}\left(\frac{R_0}{R(t)}\right)^4\left(e^{t/t_{Sal}}-1-\frac{t}{t_{Sal}}\right) .
\label{vanal}
\eeq

From Eq.~\ref{vanal} it can be seen that at late times ($t>>t_{Sal}$) the bubble velocity grows exponentially according to $v(t)\sim R_0/(5t_{Sal})\, e^{t/(5t_{Sal})}$, i.e. it approaches the derivative of Eq.~\ref{exp}.
Instead, at early times ($t<<t_{Sal}$), the bubble velocity follows the relation

\beq
v(t) \simeq \left(\frac{4}{33}\right)^\frac{1}{5}\frac{3}{5}\frac{R_0}{t_{Sal}}\left(\frac{t}{t_{Sal}}\right)^{-\frac{2}{5}},
\label{vanalearly}
\eeq i.e. we recover the $v(t)\propto t^{-2/5}$ dependence found by \citet{w77} and \citet{dw92} for a constant source of power (Eq.~\ref{vanalearly} can be  obtained by taking the derivative of Eq.~\ref{ww}).

As shown in Fig.~\ref{allinone}, the most notable feature in the velocity curve is that it reaches a minimum of $\sim$220 km~s$^{-1}$ at a few Salpeter times, and then starts increasing exponentially.
This means that the shell undergoes an acceleration and it becomes Rayleigh-Taylor unstable around a few Salpeter times, i.e. at 200-300 Myr after the BH starts accreting.
The shell stability is discussed in Section 5.

\begin{figure}[t]
  \includegraphics[angle=0, width=9cm]{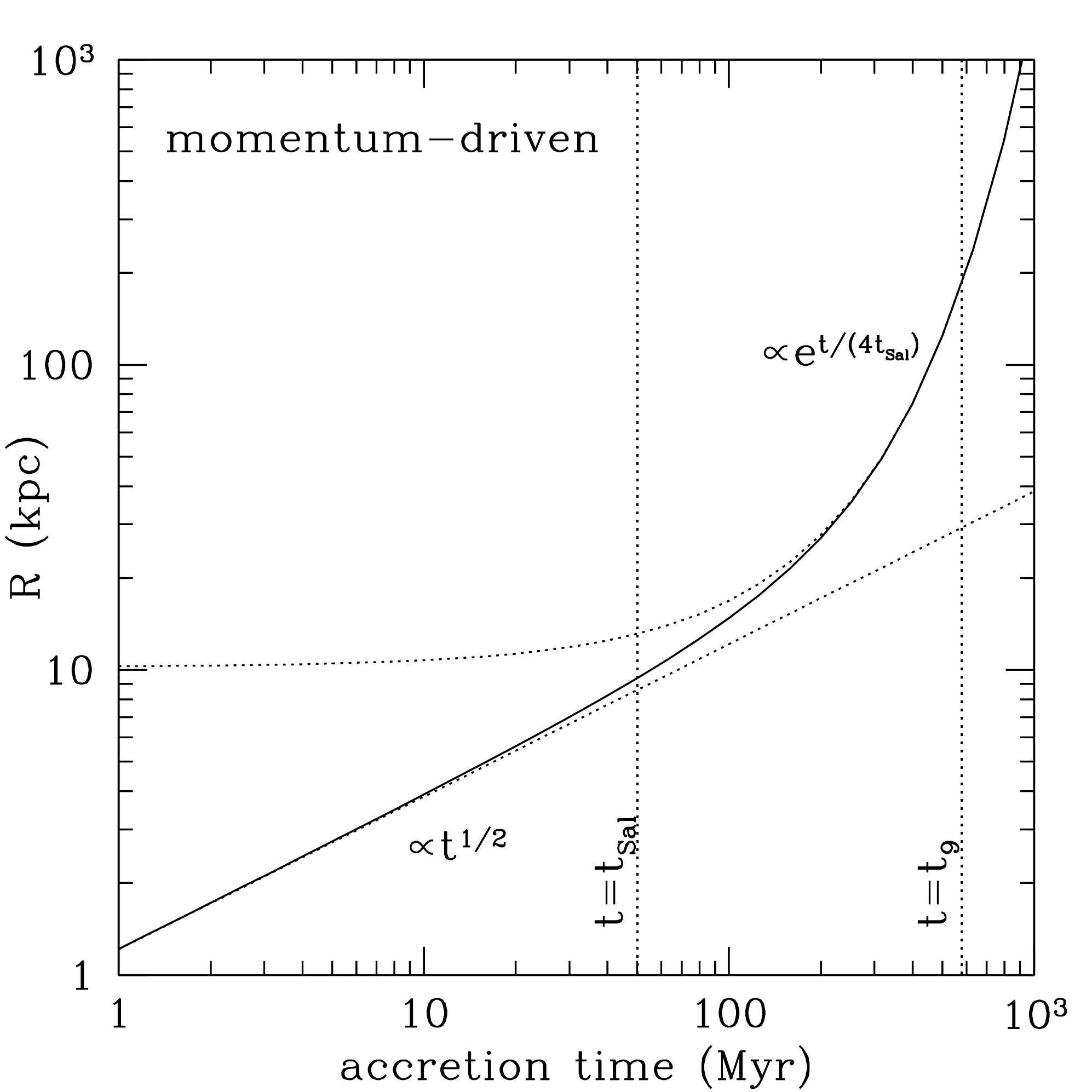}
\caption{Time evolution of the shell radius for a momentum-driven outflow produced by an early SMBH accreting at its Eddington limit.
 Input parameters are as in Fig.~\ref{rvm}. The black curve shows the analytic solution described by Eq.~\ref{analc2}.
  The vertical dotted lines are as in Fig.~\ref{rvm}.
  The two asymptotic solutions, where the shell radius scales as $t^{1/2}$ for $t<<t_{Sal}$ and $e^{t/(4t_{Sal})}$ for $t>>t_{Sal}$, are also shown as dotted lines.
} 
\label{mc}
\end{figure}

 \begin{figure}[ht]
\includegraphics[angle=0, width=7.8cm]{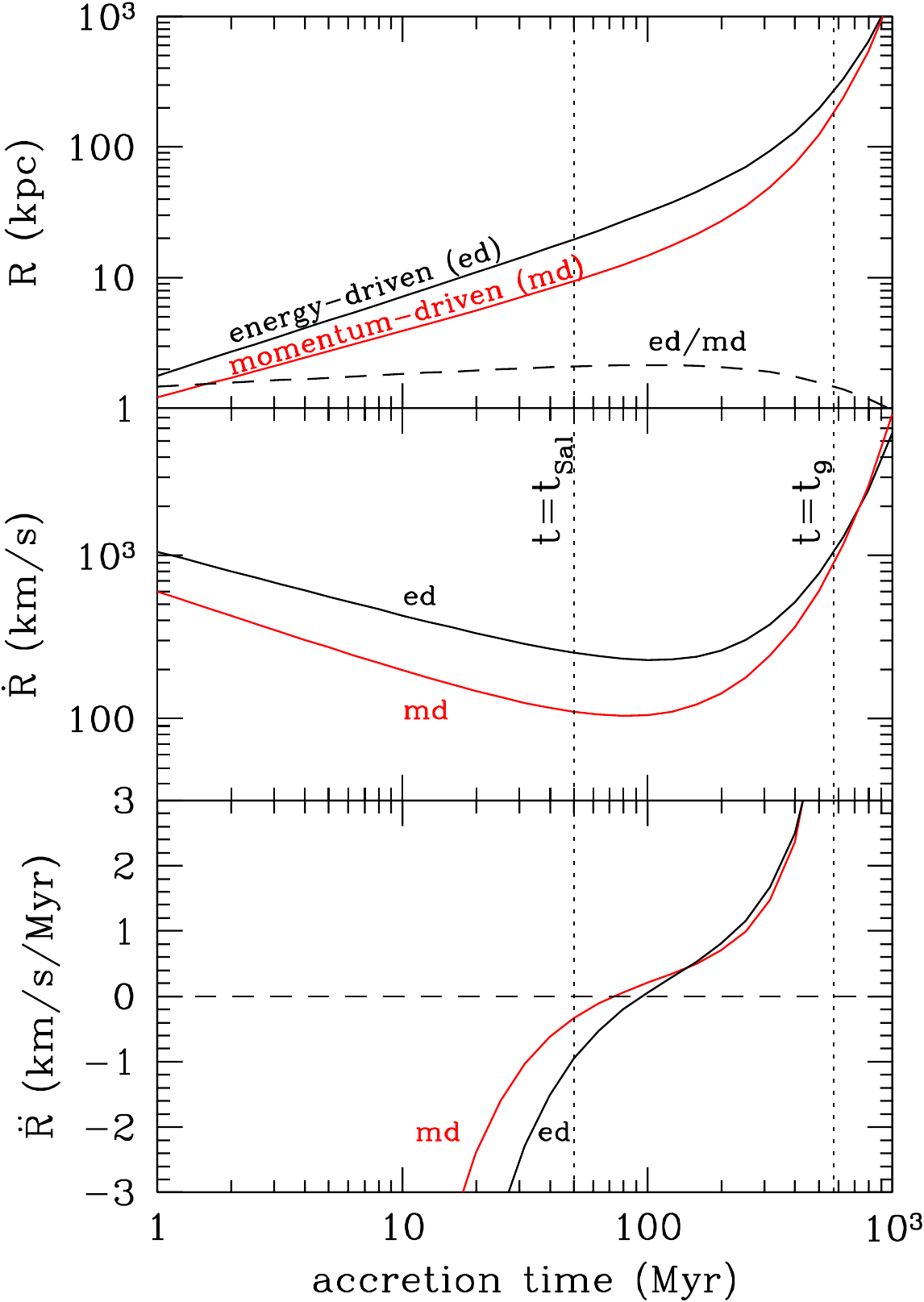}
\caption{Comparison between the time evolution of the shell radius ({\it top panel}), velocity ({\it middle}), and acceleration ({\it bottom}) for an outflow produced by an
exponentially growing black hole in the energy-driven (black curves) and momentum-driven (red curves) limits. The source input parameters are as in Figs.~\ref{rvm} and ~\ref{mc}. 
The ratio between the two radii is plotted as a dashed curve in the top panel. The vertical dotted lines are as in Fig.~\ref{rvm}. At any given time, the energy-driven outflow is about two times more
extended and faster than the momentum-driven outflow (apart from very late times, $t > 1$ Gyr, where the ejected wind mass becomes comparable with the shell mass and our treatment is inaccurate).} 
\label{cfr_radii}
\end{figure}

\subsection {Momentum driven case (maximal cooling)}

In momentum driven outflows, the shocked wind cools immediately and the bubble (region $b$ in Fig.~\ref{ofs}) collapses, so  the wind ram pressure is directly pushing on the gas shell (see e.g. Zubovas \& King 2012, Costa et al. 2014). 
By combining Eqs.~\ref{cmom} and ~\ref{ram} and recalling
that the mass of the shell $M(R)=(4\pi/3) R^3 \rho_{gas}$ for a uniform gas density $\rho_{gas}$, we obtain

\beq
\frac{d}{dt}[\frac{4\pi}{3}\rho_{gas} R^3 \dot R] = \frac{2 L_{0,w}}{v_w}e^{t/t_{Sal}} \; .
\label{analc1}
\eeq

By integrating Eq.~\ref{analc1} twice in time and solving for $R$, we obtain

\beq
R(t)=R_0 \left(e^{t/t_{s}}-1-\frac{t}{t_{s}}\right)^{\frac{1}{4}} \; ,
\label{analc2}
\eeq

where

%\begin{eqnarray}
\begin{align}
R_0 = \left(\frac{6L_{w,0}}{\pi v_w \rho_{gas} }\right)^\frac{1}{4}t_{s}^{\frac{1}{2}} \\ \notag
& \hspace{-2.5cm} = 10.5\;kpc \; \left(\frac{9\epsilon}{1-\epsilon}\right)^\frac{1}{2}\lambda^{-\frac{1}{2}}\left(\frac{f_w}{0.05}\right)^\frac{1}{4}\left(\frac{M_0}{10^4 M_{\odot}}\right)^\frac{1}{4}\left(\frac{\rho_{gas}}{10^{-28} g\;cm^{-3}}\right)^{-\frac{1}{4}}  .
\end{align}
%\end{eqnarray}

This solution is physically meaningful, as $R \rightarrow 0$ for $t \rightarrow 0$.

At early times, $t<<t_s$, Eq.~\ref{analc2} reduces to

\beq
R(t)\simeq\left(\frac{3L_{w,0}}{\pi v_w \rho_{gas}} \right)^\frac{1}{4}t^{\frac{1}{2}} \; .
\eeq

The above asymptotic solution correctly recovers the $R \propto t^{\frac{1}{2}}$ dependence found by \citet{steigman75} for momentum driven outflows
in the case of constant power sources (at early times the exponential term can  indeed be approximated with a constant).
At late times, $t>>t_s$, the exponential term takes over and the equation of motion collapses to

\beq
R(t)\simeq\left(\frac{6L_{w,0}}{\pi v_w \rho_{gas}}\right)^\frac{1}{4}t_{s}^{\frac{1}{2}}e^{t/(4t_{s})} \; .
\label{analclate}
\eeq

Eq.~\ref{analclate} shows that the e-folding time of the radius of the expanding shell is 4 times the Salpeter time, i.e. the shell expansion is
slower than that of the SMBH and of the power it releases. The expansion of the shell is shown in Fig.~\ref{mc}, where its asymptotic behaviour is also highlighted.
By taking the time derivative of Eq.~\ref{analc2} we also derived an expression for the shell velocity in the momentum driven regime as follows:

\beq
v(t)=\frac{R_0}{4t_{Sal}}\left(\frac{R_0}{R(t)}\right)^3(e^{t/t_{Sal}}-1) .
\eeq

In Fig.~\ref{cfr_radii} ($top$ and $middle$ panels) we compare the time evolution of the shell radius and velocity in the energy-driven vs. momentum-driven scenarios: 
the shell radius in the 
energy driven case is always larger -- by about a factor of two -- than in the momentum driven case, apart from very late times 
when the faster exponential growth of the momentum driven case takes over.
Similarly, the shell velocity in the energy driven regime is always higher than in the momentum driven regime except at very late times.
It is worth stressing that for $t\gtrsim1$ Gyr the mass ejected by the BH (growing as $e^{t/t_{Sal}}$) will be larger than the mass of the shell 
(which, for instance, grows as $e^{0.75 t/t_{Sal}}$ in the momentum driven regime). 
In such conditions our system of equations, which is designed to describe the motion of a dense shell enclosing an empty bubble, is no longer valid. 
In the $bottom$ panel of Fig.~\ref{cfr_radii} we also show the behaviour of the shell acceleration in the two regimes: in the energy driven limit, the acceleration
crosses zero at later times. We  return to this  in the stability analysis in Section 5.

\subsection {Effects of decreasing gas density because of cosmological expansion}

We discuss here the effects of the Hubble expansion on the evolution of the bubble radius. Since we placed the accreting BH within an unbound region, the
net effect of cosmic expansion is a reduction in the (proper) average background gas density. The relation between cosmic time and redshift for a flat cosmological
model with $\Omega_{\Lambda}\neq 0$ can be written as (e.g. \citealt{longair})

\beq
t_U(z)=\frac{2}{3H_0\Omega_{\Lambda}^{1/2}} ln \left( \frac{1+cos\theta}{sin\theta} \right), 
\label{tu}
\eeq

where  $tan(\theta)=(\Omega_m/\Omega_{\Lambda})^{1/2}(1+z)^{3/2}$. 

Eq.~\ref{tu} shows that at z=6 the age of the Universe is $t_U=914$ Myr and that for a total accretion time $t_9$ of 575 Myr, the epoch at which accretion
is supposed to start ($t_U=914-575=339$ Myr) corresponds to $z\sim12.5$.

From Eq.~\ref{rho} it is then easy to see that the average baryon density of the Universe decreases by a factor of $\sim 7$
from $z=12.5$  to $z=6$. In the energy-driven case, the normalization of the shell radius 
$R_0$ scales as $\rho_{gas}^{-1/5}$, so  at early times the bubble radius can be expected to be  a factor of  $\sim7^{0.2}$ ($\sim1.5$) smaller than that shown in Fig.~\ref{rvm}. In the momentum-driven limit, where $R_0\propto \rho_{gas}^{-1/4}$, the bubble radius would decrease by a factor of $\sim7^{0.25}$ ($\sim1.6$) at most at early times.

\subsection {Effects of the external pressure}

In Section 3 we derived the equation of motion of the shell neglecting the effects of the pressure of the ambient gas which in principle may slow down  the expansion of the bubble significantly and alter the physical properties of the shell. In particular, if the expansion velocity of the bubble is comparable to the sound speed in the ambient gas, no shock will be generated and consequently no shell of dense gas will form. In the redshift range considered here (z=6-12.5), i.e. before a full re-ionization
of the IGM, we can assume that the temperature of the IGM around a QSO is $T\approx 2\times10^4$ K at most if either the QSO radiation or that of stars in its host or in nearby galaxies is heating the IGM through photoionization \citep{ferrara00,mvw10,kaki16}. The sound speed in the IGM, $c_s = (\gamma k_B T / \mu m_p)^{1/2}$ will then be  $\sim 22$ km~s$^{-1}$ at most (assuming $\gamma=5/3$ and a mean molecular weight per particle of $\mu=0.59$, valid for primordial gas). 
This is much smaller than the minimum shell velocity computed in the previous section for  the energy-driven and the momentum-driven driven scenarios. The bubble expansion is therefore always supersonic and, as opposed to the case of a constant power source where the shell velocity decreases and the ambient pressure exerts a significant resistance at late times \citep{w77}, in the exponential case the bubble velocity grows at late times, 
making the effect of the ambient pressure negligible.  
As discussed in Section~6, the effects of the external pressure are instead significant even for an exponentially increasing source power if the QSO is placed at the centre of a large dark matter 
halo.

\subsection {Effects of gravity}

In Sections 3 and 4 we derived the equation of motion of the shell neglecting the gravitational pull exerted on it by the total mass $M_t(<R)$ enclosed within the expanding
shell radius. When including gravity, the energy and momentum equations, Eq.~\ref{cmom} and Eq.~\ref{cene}, turn into

\beq
\frac{d}{dt}[M_s(R)\dot R] = 4\pi R^2P -  \mathcal{F}_g \; ,
\label{cmomg}
\eeq

and

\beq
\dot E_{th} = L_w - 4\pi R^2 P \dot R -  \mathcal{F}_g\dot R \; ,
\label{ceneg}
\eeq respectively, where $\F_g=\frac{GM_s(R)M_t(<R)}{R^2}$ is the gravitational force exerted by the total mass $M_t(<R)$ enclosed within the radius of the shell of swept-up gas on the shell mass itself $M_s(R)$, and $\F_g\dot R$ is the rate of work done to lift the shell out of the gravitational potential. For a uniform density field $M_s(R)=(4\pi/3) R^3 \rho_{gas}$, and, for sufficiently large radii where the black hole attraction can be neglected,
$M_t(<R)=(4\pi/3) R^3 (\rho_{gas}/f_{gas})$, where $f_{gas}=\rho_{gas}/\rho_{m}$ is the gas mass fraction of the density field. By combining Eq.~\ref{ep} with Eq.~\ref{cmomg}
and Eq.~\ref{ceneg}, after some algebra we get the following equation of motion of the shell:

\beq
 15R^2\dot R^3 + 12R^3\dot R\ddot R + R^4 \dddot R + 12\pi G \frac{\rho_{gas}}{f_{gas}} R^4\dot R= \frac{3}{2\pi}\frac{L_w(t)}{\rho_{gas}}.
\label{dw92g}
\eeq

The above simple generalization of Eq.~\ref{dw92} accounts for the effects of gravity on a gas shell expanding in a constant density field.
It is indeed identical to Eq.~\ref{dw92}, but for the addition of the gravitational term $\propto R^4 \dot R$ on the left-hand side.

We  explored under which conditions the pull exerted by gravity significantly slows down the shell expansion  compared to  what was  shown in Section 4.
To make a comparison with the original solutions of \citet{w77} and \citet{dw92}, we first considered a constant energy source $L_w(t)=L_{w,0}$, which, in the absence of gravity,
produces a bubble expanding as $R(t)\propto t^\frac{3}{5}$. We then tried a power-law solution to Eq.~\ref{dw92g} of the form $R(t)=R_0t^\alpha$, which leads to the
relationship

\begin{align}
R_0^5 t^{5\alpha-3}\left[15\alpha^3+12\alpha^2(\alpha-1)+\alpha(\alpha-1)(\alpha-2) + 12\alpha \pi G \frac{\rho_{gas}}{f_{gas}}t^2\right] \\ \notag
& \hspace{-8cm} = \frac{3}{2\pi}\frac{L_{w,0}}{\rho_{gas}} \; .
\end{align}

For $\alpha=\frac{3}{5}$ the above equation becomes

\beq
R_0^5\left[\frac{231}{125}+ \frac{36}{5}\pi G \frac{\rho_{gas}}{f_{gas}}t^2\right] =  \frac{3}{2\pi}\frac{L_{w,0}}{\rho_{gas}} \; ,
\eeq

which, for $t\rightarrow 0$, returns $R_0=\left(\frac{125}{154\pi}\frac{L_{w,0}}{\rho_{gas}}\right)^\frac{1}{5}\equiv R_{early,0}$, i.e. at early times the solution of \citet{w77} applies:

\beq
R_{early}(t)=R_{early,0}\;t^\frac{3}{5} \; .
\label{early}
\eeq

Instead, for $\alpha=\frac{1}{5}$ the equation can be written as

\beq
R_0^5\left[\frac{3}{125}t^{-2}+ \frac{12}{5}\pi G \frac{\rho_{gas}}{f_{gas}}\right] = \frac{3}{2\pi}\frac{L_{w,0}}{\rho_{gas}} \; ,
\eeq

which, for $t\rightarrow \infty$, returns $R_0=\left(\frac{5}{8\pi^2 G}\frac{L_{w,0} f_{gas}}{\rho_{gas}^2}\right)^\frac{1}{5}\equiv R_{late,0}$, and the shell equation of motion at late times becomes

\beq
R_{late}(t)=R_{late,0}\;t^\frac{1}{5} .
\label{late}
\eeq

Therefore, at late times, when the enclosed mass becomes sufficiently large, gravity is effective, and the bubble radius $R(t)$ expands at a lower rate ($\propto t^\frac{1}{5}$)
than in the standard no-gravity case of \citet{dw92} ($\propto t^\frac{3}{5}$).

A characteristic transition time $t_c$ between the two regimes can be defined as the time where the two asymptotic limits on the bubble radius
described by Eq.~\ref{early} and Eq.~\ref{late} cross each other, i.e. $R_{early}(t_c)=R_{late}(t_c)$. This returns

\beq
t_c=\left(\frac{77}{100\pi}\frac{f_{gas}}{\rho_{gas}}\right)^\frac{1}{2} = 2.43 Gyr \left(\frac{f_{gas}}{0.16}\right)^\frac{1}{2} \left(\frac{\rho_{gas}}{10^{-28}g\;cm^{-3}}\right)^{-\frac{1}{2}} \; .
\label{tcrit}
\eeq

\begin{figure}[h]
\includegraphics[angle=0, width=9cm]{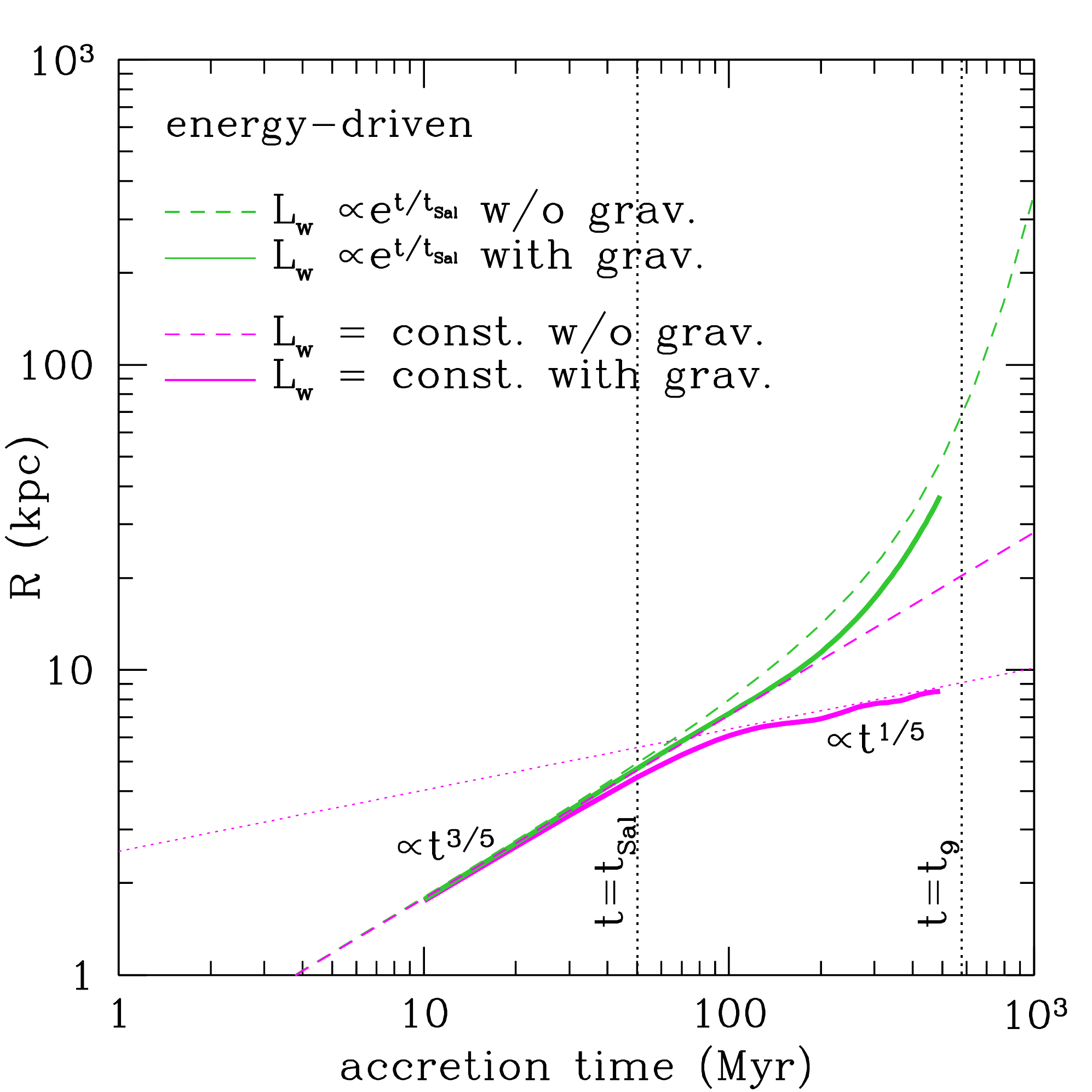}
\caption{Time evolution of the bubble radius for a constant (magenta curves) and exponentially growing (green curves) source in the energy-driven limit.
  The solid (dashed) lines show the radius evolution considering (neglecting) the effects of gravity. Solid curves have been obtained
  by numerically solving Eq.~\ref{dw92g}. The assumed parameters for the exponential input energy source $L_{w}(t)=L_{w,0}e^{t/t_{Sal}}$
  are the same as those in Fig.~\ref{rvm}. The constant source has $L_{w}(t)=L_{w,0}$. The vertical dotted lines are as in Fig.~\ref{rvm}. The density of the ambient medium
  is assumed to be $\rho_{gas}=10^{-25}$g~cm$^{-3}$, i.e. about 1000 times the average of the Universe at z=6. The power-law asymptotic behaviour  at
  early ($R\propto t^{3/5}$, magenta dashed line) and at late ($R\propto t^{1/5}$, magenta dotted line) times can be appreciated for the constant energy source.
  For the exponentially growing case, such a density field reduces the size of the bubble radius by $\sim25$\% at late times.
} 
\label{wg}
\end{figure}

From Eq.~\ref{tcrit} it is easy to see that gravity effects are negligible for a bubble inflated by a QSO expanding in a region with density equal to the mean matter density
of the Universe at z=6 ($\rho_{gas}=10^{-28}$g~cm$^{-3}$). Instead, if the expansion happens within highly overdense regions, or at very early times when the Universe was smaller and denser, gravity effects may become relevant. In Fig.~\ref{wg} we show the dragging effect exerted by a density
field with $\rho_{gas}=10^{-25}$g~cm$^{-3}$ on the bubble radius produced by a constant source with $L_{w}(t) = L_{w,0}=6.3\times10^{40}$ erg~s$^{-1}$ (the full solution to Eq.~\ref{dw92g} was obtained numerically).
In this case, $t_c\sim 77$~Myr, and the asymptotic behaviour of the bubble radius both at early and late times can be appreciated.

We then considered the case of an exponentially growing source of energy $L_{w}(t) = L_{w,0} \;e^{t/t_{Sal}}$.
Similarly to Eq.~\ref{dw92}, even Eq.~\ref{dw92g} admits a simple analytic (exponential) solution of the form
$R(t) =  R^G_0 \; e^{t/(5t_{Sal})}$, which differs from
that obtained in the case of no gravity (Eq.~\ref{exp}) only by its normalization. 
The relation between the two normalizations $R^G_0$ and $R_0$ (Eq.~\ref{r0}) is 

\beq
R^G_0 = R_0 \left(1 + \frac{\rho_{gas}}{\rho^G_{gas}}\right)^{-\frac{1}{5}},
\eeq

where $\rho^G_{gas}\equiv \frac{7}{75\pi G}t_{Sal}^{-2}f_{gas}$. For $\rho_{gas}<<\rho^G_{gas}$, $R^G_0 \sim R_0$, and the effects of gravity are negligible.
Assuming a universal gas fraction of $f_{gas}=0.16$ and the same
Salpeter time used in the previous section ($t_{Sal}=$50 Myr), we get $\rho^G_{gas}=3\times10^{-26}$~g~cm$^{-3}$. Therefore, in the case of a QSO placed within a medium
with density similar to the average at z=6, $\rho_{gas}=10^{-28}$ g cm $^{-3}$, the effects of gravity can safely be neglected. 
We note that, because of the weak dependence of the bubble radius on the density ($R\approx \rho_g^{-0.2}$), even by assuming a three dex higher background density,
one would get $R^G_0 \sim 0.75 R_0$; in other words,  for an exponentially growing source of power the gravity would reduce the bubble radius by only 25\% at late times
(see the numerical solutions in Fig.~\ref{wg}).

\begin{figure}[t]
  \includegraphics[angle=0, width=9cm]{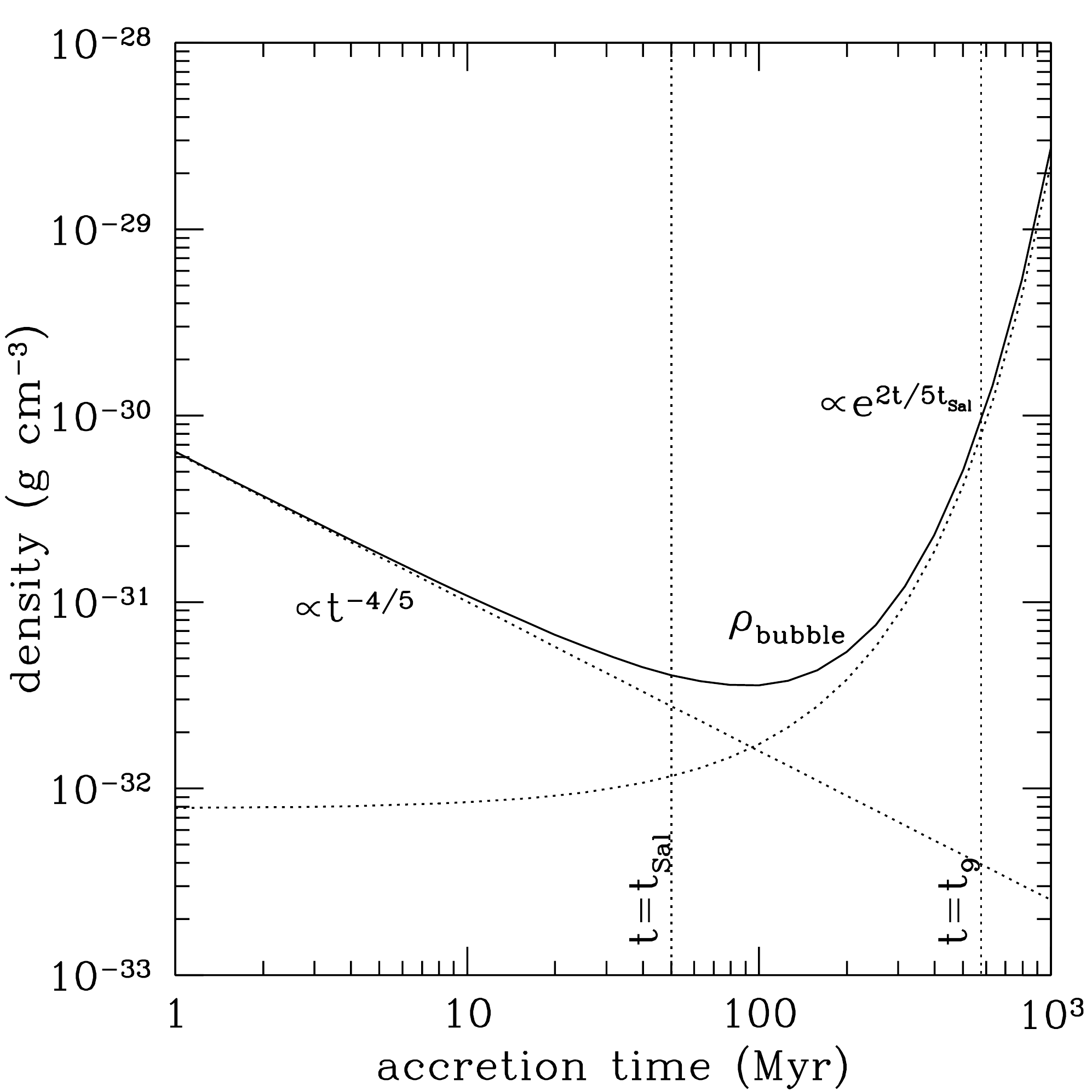}
\caption{Time evolution of the gas density in the bubble $\rho_{bubble}$ for the energy-driven case (region $b$ in Fig.~\ref{ofs}). The source input parameters are as in Fig.~\ref{rvm}. The black curve shows the analytic solution described by Eq.~\ref{denseq}.
  The vertical dotted lines are as in Fig.~\ref{rvm}.
  The two asymptotic solutions, where the gas density scales as $t^{-4/5}$ for $t<<t_{Sal}$ and $e^{2t/(5t_{Sal})}$ for $t>>t_{Sal}$, are also shown as dotted lines.
} 
\label{dens}
\end{figure}

\section {Shell stability}

The basic structure of wind-driven bubbles was given in Section~2. We now  analyse the stability of the shells. For our terminology we refer to 
to the ambient medium that has passed through the outermost bubble shock as the ``shell'' (region $c$ in Fig~\ref{ofs}). The wind material that has passed through the reverse shock, where the driving wind from the central black hole and its accretion disc impinges on the bubble, is called the shocked wind medium or simply the ``bubble'' (region $b$ in Fig~\ref{ofs}). There are two main instabilities:
\begin{enumerate}
\item The Rayleigh-Taylor instability, which has a classic growth rate given by $\sqrt{2\pi a/\lambda}$ , where $a$ is the acceleration and $\lambda$ is the spatial scale, and where the classic instability pattern of bubble and spikes occurs as a result of the instability. This is highly disruptive for shells in the linear analysis.

\item The Vishniac instability \citep{vishniac83}, which occurs for thin shells bounded by thermal pressure on one side and ram pressure on the other side. If the shell is perturbed, the ram pressure has an oblique component
on it, whereas the thermal pressure is always normal to the shell surface. As a result, material is transported along the shell causing it to oscillate and producing an ``overstability''.
\end{enumerate} 
In the initial phases, i.e. at $t<100$ Myr, when the bubble growth is similar to that of a constant energy source, the decelerated shell may indeed be prone to the Vishniac overstability \citep{pittard13,krause14}. 
However, numerical results in the non-linear regime have shown that these oscillations saturate and do not lead to any real disruption \citep{maclow93,michaut12}.
We  therefore only discuss the Rayleigh-Taylor instability in the following, noting clearly that the final non-linear state will have to be calculated by high-resolution simulations.

%The difference between the classic Rayleigh-Taylor instability and the case considered here is that the detailed solutions already studied do not formally apply to the exponential propagating wind. 
%The standard Rayleigh-Taylor problem has a constant background acceleration and the careful analysis following a time dependent background \citet{bernstein1978} assumes a similarity solution $R \sim t^{\beta}$. 
%In this study, for simplicity, we assume that the time scale on which the instability occurs is short compared to the Salpeter time. In that approximation there are four cases:
 
 %\begin{enumerate}
 %\item Decelerating shell with $\rho_{shell} > \rho_{bubble}$. This is stable.
 %\item Decelerating shell with $\rho_{shell} < \rho_{bubble}$. This is unstable. 
 %\item Accelerating shell with $\rho_{shell} > \rho_{bubble}$. This is unstable.
 %\item Accelerating shell with $\rho_{shell} < \rho_{bubble}$. This is stable.

%\end{enumerate} 

Our case is different from the classic behaviour of wind-blown bubbles as the gas density within the bubble $\rho_{bubble}$ is expected 
to increase with time once the exponential term dominates, whereas the density of bubbles inflated by constant winds is expected to decrease
with time. For instance, for the stellar winds studied by \citet{w77}, $\dot m = const.$, $m \propto t$ and $R\propto t^{3/5}$, from which it follows that
$\rho_{bubble} \propto m/ R^3 \propto t^{-4/5}$.

To compute $\rho_{bubble}$ we  assume that the wind shock boundary $R_w$ (i.e. the boundary between region a and b in Fig.1) is proportional to the shell radius $R$ so that the shells move together as a piston with $R_w= \eta R$ with $\eta$ a slowly varying constant of the order of unity. 
We then obtain a simple formula for the bubble density, assuming for simplicity that this is uniform across the bubble: 

 \beq
\rho_{bubble}=\frac{\dot m_{w,0} t_{Sal} (e^{t/t_{Sal}}-1)}{(4/3)\pi R(t)^3 (1-\eta^3)} , 
\label{denseq}
 \eeq

where the numerator (ejected wind mass) has been taken from Eq.~\ref{ewm}. The time evolution of $R$ is given by Eq.~\ref{anal}.

Assuming $\eta<<1$, and considering Eqs.~\ref{exp} and ~\ref{ww}, Eq.~\ref{denseq} reduces to
$\rho_{bubble}^{late} \simeq \rho_o e^{\frac{2}{5}\frac{t}{t_{Sal}}}$ at late times ($t>>t_{Sal}$), and to
$\rho_{bubble}^{early} \simeq \rho_o (33/4)^{3/5}(t/t_{Sal})^{-4/5}$
at early times ($t<<t_{Sal}$). Here $\rho_o=\frac{3}{4\pi}\frac{\dot m_{w,0} t_{Sal}}{R_0^3}$.

The full behaviour of the bubble density and its asymptotic limits are shown in Fig.~\ref{dens}. For the parameters assumed here, the density in the bubble is always much lower than that 
in the shell $\rho_{shell} = (1+\mathcal{M}^2) \rho_{gas}$, where $\mathcal{M}$ is the Mach number. For instance, the minimum shell velocity of $\sim220$ km~s$^{-1}$ in the energy-driven case
considered here (see central panel of Fig.~\ref{cfr_radii}) would correspond to a minimum Mach number of $\sim 10$ and hence to a minimum shell density of $\rho_{shell}\sim10^{-26}$ g~cm$^{-3}$. At early times, when the shell is decelerating, the bubble density follows
the $\propto t^{-4/5}$ behaviour expected for winds outflowing at a constant mass rate. In this regime, the shell is Rayleigh-Taylor stable. At later times ($t>100$ Myr), when the 
exponential term dominates, the shell accelerates and is expected to be Rayleigh-Taylor unstable. As shown in Fig.~\ref{cfr_radii} (bottom panel), in the energy-driven limit, 
the shell acceleration crosses zero at $t\sim 2t_{Sal}$, i.e. this is the time where Rayleigh-Taylor instabilities start developing. Interestingly,
for momentum-driven outflows, this instability time is shorter as the acceleration crosses zero at about 1.4$t_{Sal}$.
We note that in the fully exponential regime the ejected mass grows more rapidly than the bubble volume, producing an increasing gas density within the bubble. However,
only for accretion times longer than 1 Gyr (i.e. BH masses $ >>10^{10} \; M_{\odot}$) or large values of $\eta$ ($\eta>0.9$, i.e. the QSO wind freely streaming up to
hundred kpc distances) $\rho_{bubble}$ might be larger than $\rho_{shell}$; however, these two situations appear
 unphysical. Furthermore, as specified in the previous section, for $t>1$ Gyr, the ejected mass becomes comparable with the shell mass and our set of equations is no longer valid.

\begin{figure}[t]
\includegraphics[angle=0, width=9cm]{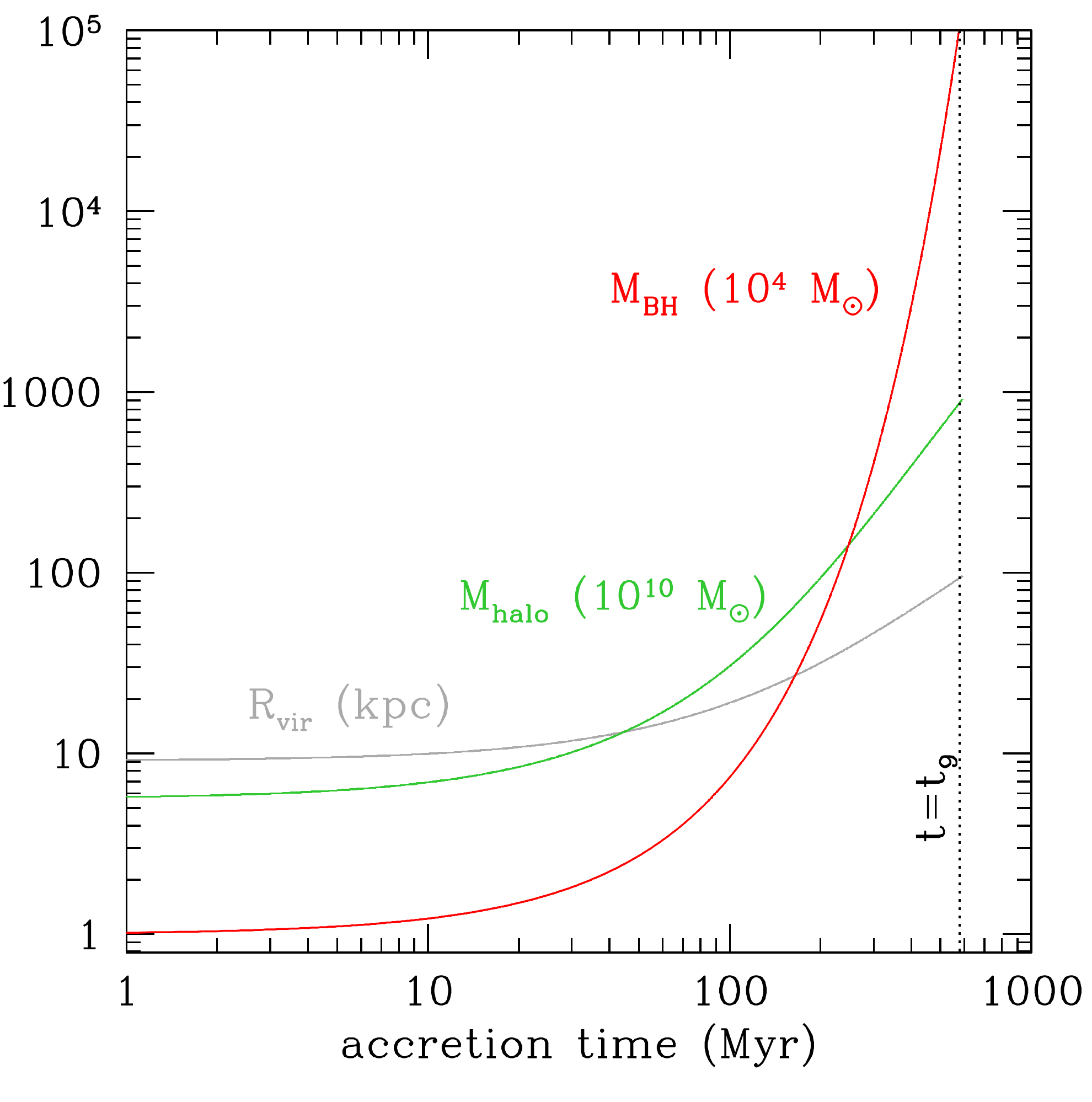}
\caption{Time evolution of the halo mass (in units of $10^{10}\;M_{\odot}$, green curve) and virial radius (in units of kpc; grey curve) as compared to the evolution 
of the black hole mass  (in units of $10^4\;M_{\odot}$; red curve) during the QSO accretion history. By the time the SMBH has grown to $10^9\;M_{\odot}$, the halo mass 
and virial radius have grown by a factor of $\sim150$ and $\sim10$, respectively.} 
\label{masses}
\end{figure}

\begin{figure}[t]
\includegraphics[angle=0, width=9cm]{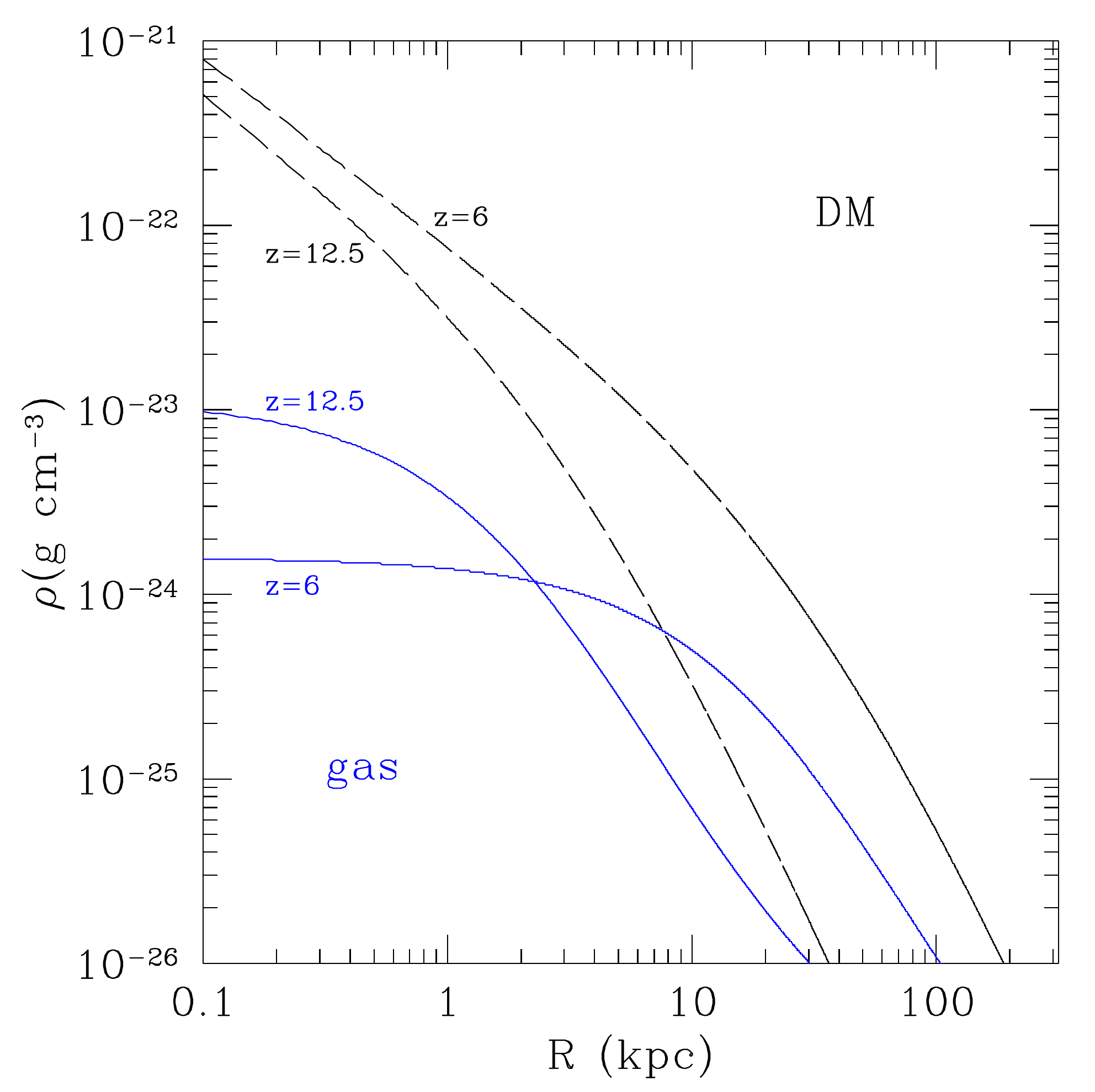}
\caption{Density profiles of dark matter (black dashed curves) and gas (solid blue curves)  at z=12.5 and z=6 (as labelled) for a QSO hosting halo growing to $10^{13}\;M_{\odot}$ by z=6.
The gas density is up to five orders of magnitude higher than in the uniform density field case explored in Section 4.} 
\label{prof}
\end{figure}

\begin{figure}[t]
\includegraphics[angle=0, width=9cm]{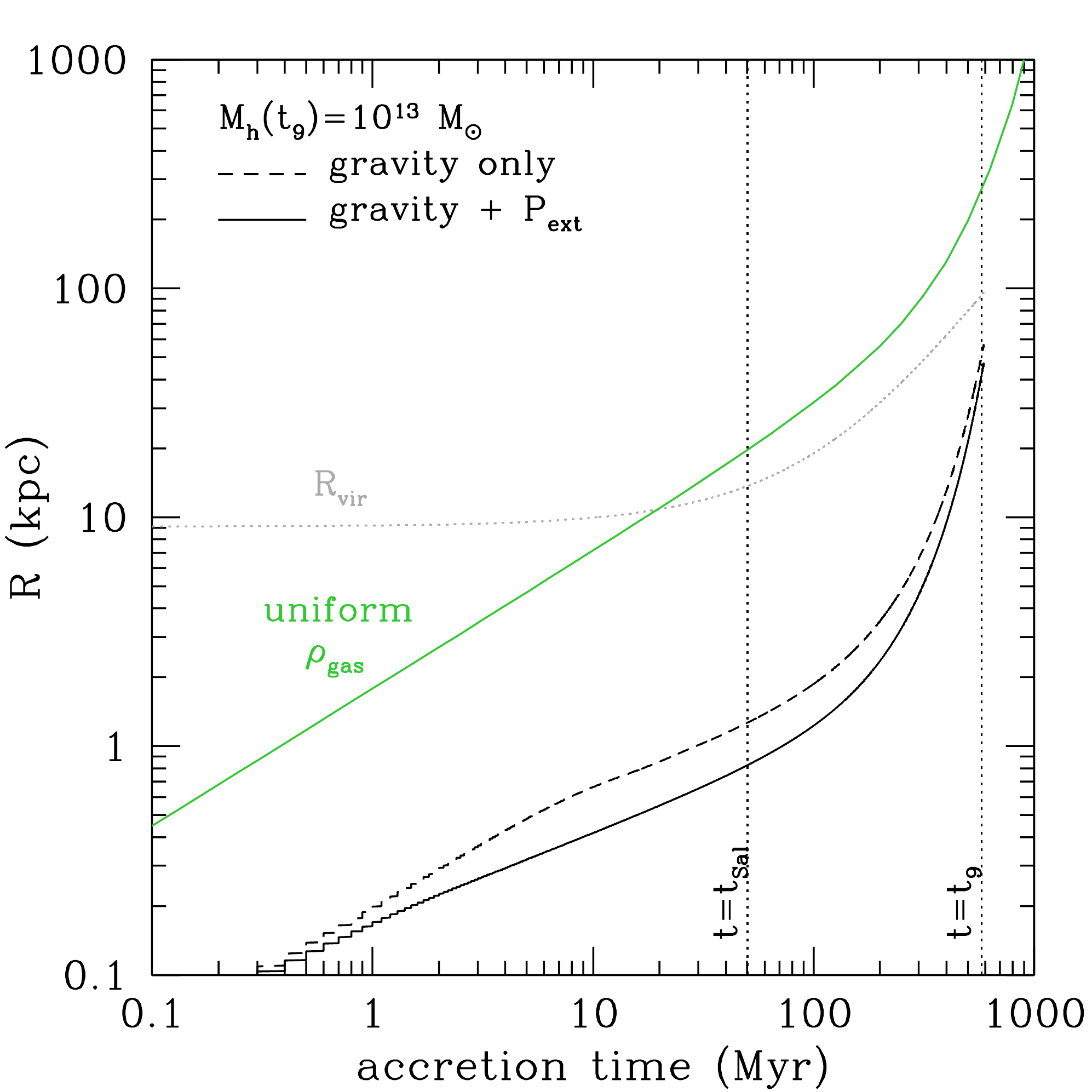}
\caption{Time evolution of the bubble radius produced by a growing SMBH in a dark matter halo growing to $10^{13}\;M_{\odot}$ at z=6 
in the energy-driven case (black curves). The dashed curve shows the evolution of
the bubble radius when only the gravitational pull of the halo is considered. The solid curve also considers  the external pressure $P_{ext}$ by the halo gas. 
The mass and luminosity of the accreting SMBH are as in Fig.~\ref{rvm}. The evolution of the halo virial radius is also shown (grey curve).
For comparison, the evolution of the bubble radius in the case of a uniform density field as assumed
in Fig.~\ref{rvm} is also plotted (green curve). 
The vertical dotted lines are as in Fig.~\ref{rvm}.} 
\label{rhalo}
\end{figure}

\section {QSO in a growing dark matter halo} 

We explore here the effects of placing the accreting BH within a spherically symmetric massive dark matter halo. In particular, we follow the evolution of the bubble as the dark matter
halo grows between z=12.5 and z=6 and so does its gas content. In fact, the mass of early dark matter halos can grow by
more than two orders of magnitude in this redshift range (see e.g. \citealt{correa15}). Furthermore, the gas content of the halo is expected to change its radial profile
(e.g. \citealt{fl13}). All these effects, which are generally neglected in simple analytic models, regulate the expansion rate of the outflows produced by early QSOs. 

The typical mass of the halos in which $z\sim6$ QSOs reside is largely unknown. Based on abundance-matching arguments, early SMBHs should live in
$10^{13}\;M_{\odot}$ halos if their duty cycle is of the order of unity. On the other hand, for smaller duty cycles, or if a large, hidden population of early obscured QSOs exists (which is likely, based
on what is known at lower redshifts), early SMBHs should be hosted by smaller, more abundant halos. Furthermore, some simulations suggest that QSO feedback may preferentially 
inhibit gas accretion in large halos, and that early SMBHs at $z\sim6$ are hosted by halos with $M_h\sim3\times 10^{11} M_{\odot}$ on average \citep{fanidakis13}.
 
To trace the evolution of massive DMHs at early times up to the largest mass scales, we considered the results of the Millennium XXL simulation (MXXL, \citealt{angulo12}). 
The MXXL is indeed the only N-body simulation encompassing a sufficiently large volume ($\sim 70$ Gpc$^3$) to be able to follow the
growth of the rarest and most massive halos at any time. 

\citet{angulo12} report an analytic fit to the average evolution of the most massive halos from z=6 to z=0. We extrapolated that fit back in time 
by considering the average halo accretion rates as quoted in \citet{angulo12} and derived the following expression for the time evolution of massive dark matter halos from z=12.5 to z=0:
$M_h(z) = M_h(0) e^{-0.783z}$ , where $M_h(0)$ is the halo mass at redshift zero. It is then easy to see that, on average, a halo of $\sim10^{13}\;M_{\odot}$ at z=6 had a mass of $\sim6\times10^{10}\;M_{\odot}$ at z=12.5 and should evolve
into a $10^{15}\;M_{\odot}$ halo at z=0 (i.e. into a Coma-like massive galaxy cluster). We define the halo mass and virial radius in the standard way, i.e.
$M_h = (4/3) \pi R_{vir}^3 200\rho_{crit}$, where $\rho_{crit}=3H^2(z)/(8\pi G)$ is the critical density of the Universe.
The evolution of the halo mass and virial radius  compared to the evolution of the black hole mass are shown in Fig.~\ref{masses}.

We assume that, at any redshift, the dark matter density in the halo follows the Navarro-Frenk-White profile \citep{nfw97} and that as the halo grows its concentration 
follows the concentration-mass relation derived by \citet{correa15} (see their Eq.20). For the massive halos in the redshift range z=6-12.5 considered here, 
the concentration parameter is nearly constant, $c\sim2.8$.

For the gas within the halo, we consider the radial profile reported by \citet{fl13}, who extended the work by Makino et al. (1998)
on the profile of gas in hydrostatic equilibrium within halos with NFW dark matter density profiles, and allowed for gas temperatures lower than the virial temperature
$T_{vir}=(\mu m_p/2k_B)\times(GM_h/R_{vir})$. Indeed, it has been suggested that gas accreting onto dark matter halos may not be shock heated to $T_{vir}$ if the halo is smaller than a critical mass $M_{shock}$ ($\sim10^{12}M_{\odot}$ at all redshifts; \citealt{db06}). At high redshifts most halos have masses below $M_{shock}$, hence gas accretion
onto them might occur primarily through streams of gas that remain cold, and a large fraction of cold gas may be even present in larger halos (see e.g. Fig. 11 in \citealt{overzier16}). 

\begin{figure*}[t]
\begin{center}
\includegraphics[angle=0, width=13.cm]{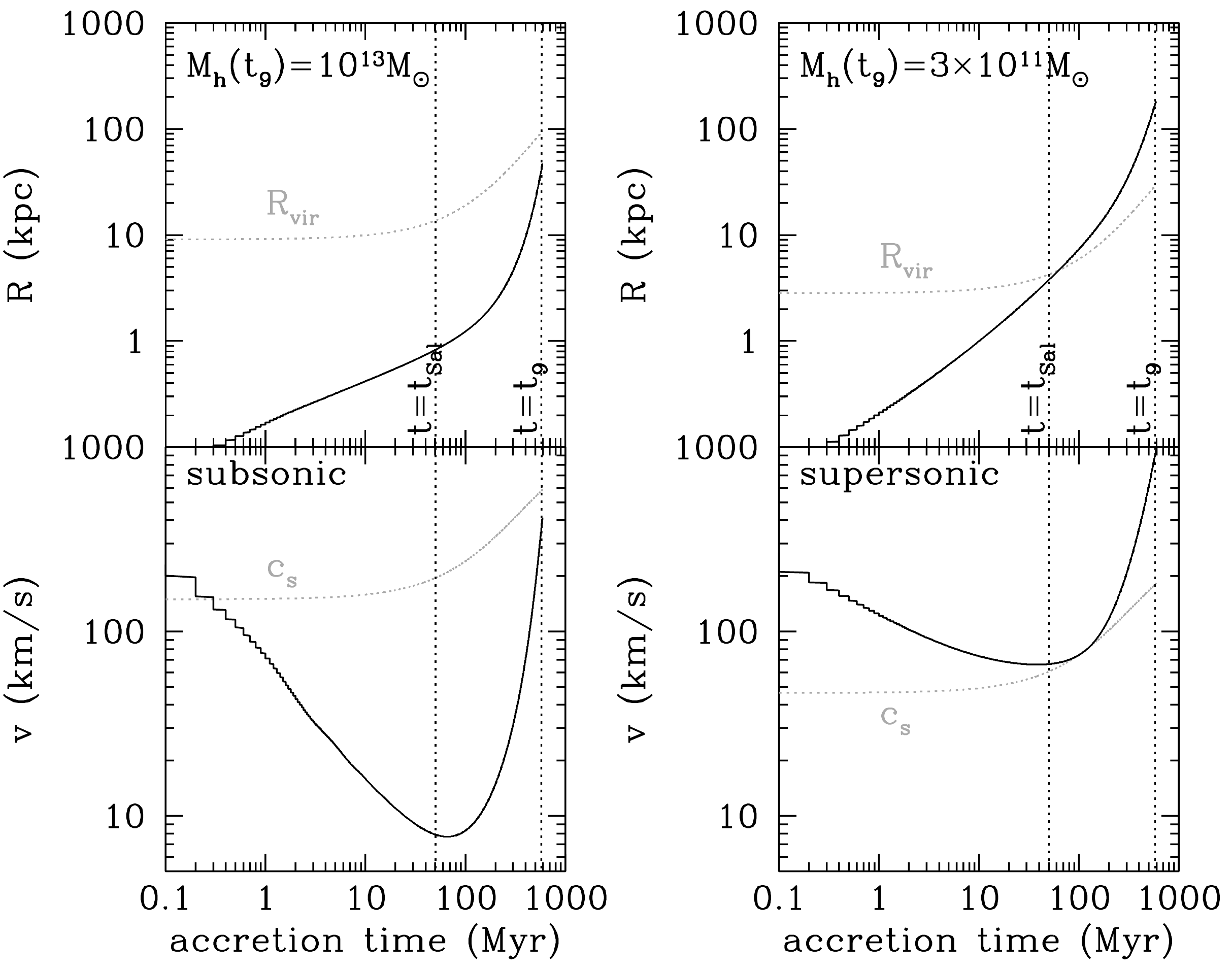}
\caption{{\it Left column:} Time evolution of the bubble radius ({\it upper panel}) and velocity ({\it lower panel}) produced by a growing SMBH in a dark matter halo growing to $10^{13}\;M_{\odot}$ by z=6 
in the energy-driven case (black curves) as compared with the evolution of the halo virial radius and sound speed of the halo gas, respectively (grey curves). For such a massive halo the outflow expansion
is always subsonic. {\it Right column:} As in the {\it left column} but
for a smaller halo that grows to $3\times10^{11}\;M_{\odot}$ by z=6. In this case the outflow expansion is always supersonic. The meaning of the vertical dotted lines is as in the previous figures.} 
\label{quad}
\end{center}
\end{figure*}

We investigated the energy driven case and modified Eqs.~\ref{cmomg} and \ref{ceneg} to account for the time evolution of the mass and profile of the dark matter and gas.
We first considered a QSO within a halo whose mass reaches $10^{13}\;M_{\odot}$ at z=6, and assumed that the ambient gas is always at $T_{gas}=T_{vir}$  as the halo grows.
In Fig.~\ref{prof} we show the halo DM and gas profiles at z=12.5 and at z=6. The QSO outflow now has  to sweep gas densities that are up to five orders of magnitude higher than in the uniform density field case explored in the previous sections.

We assumed for the accreting black hole the same input parameters as in the previous Sections (i.e. $M_0= 10^4\; M_{\odot}$, $L_{w0}=6.3\times 10^{40}$ erg~s$^{-1}$). 
The expansion of the bubble radius was obtained numerically and is shown by the dashed line in Fig.~\ref{rhalo}. Based on the results obtained in the previous  sections we are now in a position to 
understand all the phases of the bubble expansion. At early times / small radii (e.g. $<$1 Myr, or $<0.2$ kpc) the exponential source power can be approximated as a 
constant and the gradient in the gas profile is mild (see Fig.~\ref{prof}); therefore, the bubble radius expands following the classic $t^{3/5}$ law valid for constant sources within a uniform
density field. At $t\sim 10$ Myr, the source power can still be approximated as a constant and the mass enclosed within the bubble radius has become sufficiently large to slow down the expansion of the shell of swept-up gas, which now follows the $t^{1/5}$ law found in Section~4.5. Finally, for $t\gtrsim t_{Sal}$, the exponential source power takes over gravity,
and the expansion of the bubble is also exponential. At $t_9$ the bubble radius is 60 kpc. That is, the outflow has reached scales comparable with those observed
in the $z=6.4$ QSO SDSS~J1148 \citep{cicone15}. As expected, the bubble size is smaller (by a factor of $\sim4-5$) than what was
found in the case of a uniform density field since now the outflow is crossing a gas density that is on average a few dex higher. As for the velocity of the bubble expansion, this reaches $v=450$ km~s$^{-1}$ 
at $t=t_9$, whereas velocities up to 1000 km~s$^{-1}$ have been observed for SDSS~J1148. 

The computations above neglect the effects of the external pressure on the bubble evolution, which, unlike the case of a QSO placed in the field, 
are now expected to be relevant. We then modified the momentum equation (Eq.~\ref{cmomg}) by adding on the left-hand side a term equal to $-4\pi R^2 P_{ext}$, where $P_{ext}=\rho_{gas}k_BT_{gas}/\mu m_p$. As shown by the solid curve in Fig.\ref{rhalo}, the external pressure reduces the bubble radius by a factor of $\sim1.6$ up to a few Salpeter times ($\sim1.2$ at late times, in the fully exponential regime of the QSO power).  

For a given QSO power, the evolution of the bubble expansion strongly depends on the halo mass in which the BH resides. In the case of a very massive
halo ($M(t_9)=10^{13}\;M_{\odot}$), the bubble reaches radii of $\sim 50$ kpc by z=6, with velocities of $\sim 400$ km~s$^{-1}$. This is shown in Fig.~\ref{quad} (left), where the
comparison between the velocity of the bubble expansion and the sound speed of the gas in the halo is also shown. For massive halos, 
the bubble expansion is subsonic during the entire accretion history of the black hole. This means that the QSO wind is not able to create a shock and then no shell of
dense gas  propagates outwards. In this case the treatment provided by our equations can still be used to approximately describe the expansion of the contact discontinuity
between the bubble and the ambient gas \citep{koo92}. In smaller halos, the gas density encountered by the outflow is on average smaller, hence the outflow velocity is higher, 
and the sound speed in the ambient gas is lower ($c_s\propto M_h^{1/3}$). For halos with $M(t_9)\leq 3\times10^{11}\;M_{\odot}$,
the expansion turns supersonic at all times (see Fig.~\ref{quad} right) and our treatment provides a careful description of the shell motion.

\section{Discussion}

\subsection{Continuos vs. intermittent accretion}

The basic assumption in our computations is that the outflow is powered by a black hole that has been accreting mass without interruption
for many e-folding times, corresponding to $\sim580$ Myr, in the case presented before. This is necessary to produce the $10^9\;M_{\odot}$ black holes observed in
z = 6 QSOs if the accretion rate is limited to the Eddington rate\footnote{Defined as the critical accretion rate producing an Eddington luminosity for a radiatively efficient 
thin disc around a non-rotating black hole, i.e. $\dot m_E =17L_E/c^2$.}
and if the radiative efficiency is that of standard geometrically thin, optically thick Shakura-Sunyaev accretion discs, i.e. $\epsilon\sim0.06-0.3$. Recent works have shown
that in fact most QSOs at z = 6 may grow in this way given their measured radiative efficiencies and Eddington
ratios \citep{trak17}. However, the likelihood of such a long, uninterrupted accretion at high rates is highly debated. Hydrodynamical
simulations \citep{ciotti07,dubois13} have shown that strong feedback effects occur already above a fraction of the Eddington ratio
(although this is based on spherical rather than  disc accretion models). These feedback effects rapidly shut down
the accretion flow, which can possibly restart after a quiescence time longer than the burst of activity. The accretion
and shut-down episodes can continue for many cycles, but it is clear that low duty cycles make the build-up
of early SMBHs extremely challenging. A possible solution has been proposed by considering that when the accretion
rate is close to Eddington, the thin-disc solution  no longer applies, and the accretion should occur through radiatively
inefficient ``slim'' discs with $\epsilon< 0.05$ \citep{madau14,volonteri15}. We recall that the \citet{soltan82} argument indicates that the bulk of the BH growth in the Universe must occur through radiatively efficient accretion
(see also \citealt{yu02,marconi04}). However, radiatively inefficient accretion can still power sub-populations of AGN that provide little contribution to the total mass density of local SMBHs,  like e.g. QSOs at very high redshifts. 
In this case, the e-folding time of the black hole growth may
easily be much shorter than in the standard, radiatively efficient case, whereas
the radiative output would be only mildly super-Eddington \citep{madau14,volonteri15}. As an example, based on the results of \citet{madau14}, 
a non-spinning black hole that is accreting at four times its Eddington rate should have a radiative
efficiency of $\epsilon\sim0.03$ and a bolometric luminosity of $L_{bol} \sim2.6L_E$. In this case, the Salpeter time is a factor of 7 smaller
than what we assumed for continuous accretion, as is the duration of the accretion needed to grow a $10^4\;M_{\odot}$
seed to a $10^9\;M_{\odot}$ SMBH. Since the wind power is assumed to be proportional to the radiative (bolometric) QSO output,
this is 2.6 times more powerful than what we assumed for the continuous, Eddington-limited case. By recalling that
in the energy-driven limit, the normalization of the bubble radius scales approximately as $R \propto L_{w,0}^{1/5}t_{Sal}^{3/5}$, where $L_{w,0}$ is the initial wind power, 
we may expect that, in the case of Super-Eddington accretion, the bubble radius is a
factor of $2.6^{-1/5}\times7^{3/5}\sim2.6$ smaller. By considering that the bubble will keep expanding inertially during the non-active phases,
this factor is likely an upper limit. We then conclude that intermittent super-Eddington accretion should produce
bubble sizes comparable within a factor of 2 to what has been discussed in the previous sections. 
Interestingly, the low duty cycles implied by intermittent accretion would imply that early SMBHs are much more abundant than the active QSOs detected
at z=6, and then that they should be hosted on average by smaller halos which may favour the development of supersonic outflows (see next section).

\begin{figure}[t]
\includegraphics[angle=0, width=9cm]{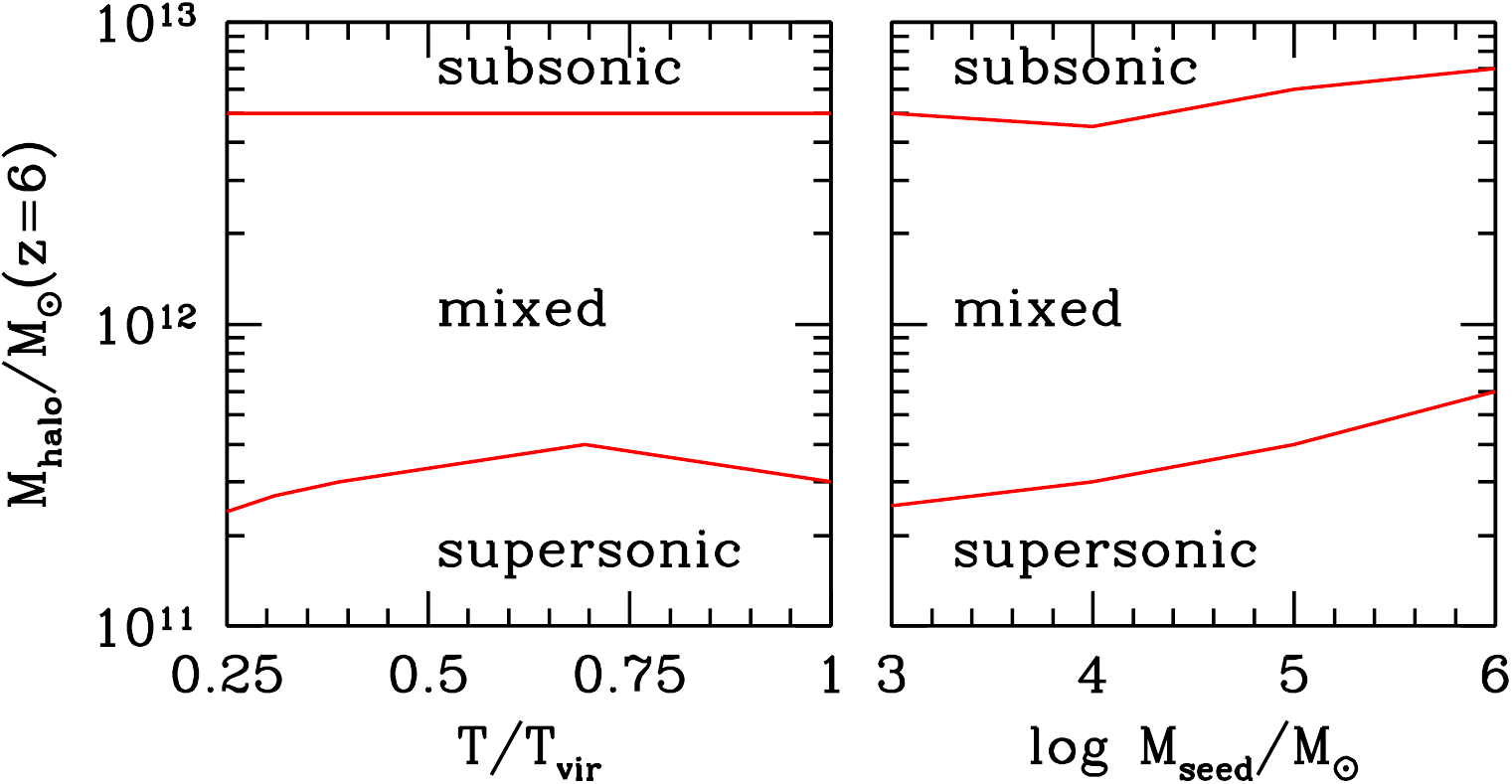}
\caption{{\it Left:} Outflow regimes as a function of halo mass at z=6 and halo gas temperature normalized to the virial temperature. 
The energy-driven limit is considered. The seed BH mass ($10^4\;M_{\odot}$) and all the other AGN input 
parameters are kept as in the previous computations. The transitions from supersonic to mixed (i.e. the outflow is part of the time supersonic and part of the time subsonic), and from mixed to 
subsonic, occur at $M_h\sim3\times 10^{11}\;M_{\odot}$ and $M_h\sim5 \times 10^{12}\;M_{\odot}$, respectively (red curves), and are nearly independent on $T_{gas}$.  {\it Right: } Same as in the {\it left} panel but as a function of halo mass at z=6 and initial seed BH mass. Here we assume $T_{gas}=T_{vir}$ for simplicity.} 
\label{grid}
\end{figure}

\subsection{Exploring the parameter space: outflow properties as a function of halo mass, gas temperature, and BH seed}

Recent hydrodynamical cosmological simulations of early BH formation suggest that part of the gas in the hosting halo may be at temperatures significantly lower than $T_{vir}$
when it is crossed by the QSO outflow \citep{dubois13}. Cold gas ($T<10^5$ K) in the form of filaments accreted by the halo itself may indeed constitute most of the gas mass in the halo. 
However, this cold gas is expected to fill only a minor fraction of the halo volume, which should  instead be mostly occupied by a hot diffuse medium with $T\sim T_{vir}$. 
We recall that $T_{vir}\propto M_h^{2/3}$. For reference, for a halo growing to $3\times 10^{11}\; M_{\odot}$ 
($10^{13}\; M_{\odot}$) by z=6, the virial temperature increases from $\sim10^5$~K ($\sim10^6$~K) at z=12.5 to $\sim1.3\times 10^6$~K ($\sim1.3\times10^7$~K) at z=6. We explored the effects of varying the gas temperature in the halo allowing the parameter $V \equiv T_{gas}/T_{vir}$ to vary in the gas profile equations of \citet{fl13}. As expected,
for $T<T_{vir}$ the gas has higher central densities and steeper profiles. 
We explored the gas temperature variations in halos
of different mass and the results are shown in Fig.~\ref{grid} (left). The seed mass of the black hole and all the other source parameters were kept as in the previous computations.
We find that the onset of the subsonic and supersonic regimes of the outflow are  nearly independent on  the gas temperature. This can be understood by considering that, on the one
hand, the sound speed decreases with decreasing gas temperature, on the other hand, the gas central density increases, slowing down the outflow expansion.
For halos with masses $M_h\lesssim3\times 10^{11}\;M_{\odot}$ the outflow is always supersonic.  As more massive
halos are considered, the outflow spends a larger fraction of time at subsonic velocities. For $M_h\geq5\times 10^{12}\;M_{\odot}$ the outflow is subsonic at any time above $t_{acc}>1$ Myr (we  simply refer to this as always subsonic).
When all the other parameters are fixed, the bubble radius and velocity of supersonic outflows can reach very large values for $T<T_{vir}$. As an example, for a halo with $M_h=3\times 10^{11}\;M_{\odot}$ 
the bubble radius and velocity at $t_9$ (z=6) are $\sim1$ Mpc and $6000$ km~s$^{-1}$ for $T\sim T_{vir}/2$, to be compared with $\sim200$ kpc and $\sim 1000$ km~s$^{-1}$ for $T=T_{vir}$. \footnote{
As noted by \citet{fl13}, the central gas densities may reach implausibly high values (and extremely steep profiles) for $T<<T_{vir}$. In our computations, this would also turn into
implausibly high values of $v(t_9)$ and $R(t_9)$, so we limited our computation to $T\geq T_{vir}/4$.}
We also investigated what happens if the seed BH mass is left free to vary in the range $10^3$ to $10^6\;M_{\odot}$ in hosting halos
whose mass may end up at z=6 in a range between $10^{11}$ and $10^{13}\;M_{\odot}$. We do not push our computations to seed masses smaller than $10^3\;M_{\odot}$ because this would mean starting
the accretion at redshifts greater than $z=17$, where i) our extrapolations for the evolution of the host halo mass are highly uncertain and ii) the central densities of these early halos as modelled here
reach implausibly high values (see \citealt{fl13}). For simplicity, we kept $T_{gas}=T_{vir}$ in the following computations. Also, we kept all the input AGN parameters
as in the previous computations ($\epsilon=0.1, \lambda=1, f_w=0.05$) so that the QSO wind power scales linearly with the seed mass. The results are shown in Fig.~\ref{grid} $(right)$.
The supersonic-to-mixed and the mixed-to-subsonic outflow transitions occur at increasing halo masses as the seed BH mass increases because varying the
seed mass means varying the initial conditions for the outflow expansion: larger seeds imply more powerful initial winds and a later start of the accretion (i.e. at a lower redshift) 
when the halo mass has grown and the central gas density is lower. This has the effect of increasing the speed of the outflow, which then needs bigger halos to be slowed down to subsonic velocities.
As for the bubble ``final'' radius and velocity at $t_9$, the same conclusions reached in Fig.~\ref{radii} apply: as the Salpeter time does not depend on the seed mass, 
the bubble radius and velocity only depend on the final BH mass, which is fixed to $10^9\;M_{\odot}$. 
Larger/smaller seeds simply mean shorter/longer accretion times to reach the same BH mass and bubble radii.

\subsection {Detection of  hot gas within QSO bubbles through the thermal Sunyaev-Zeldovich effect}

From an observational point of view, if would be difficult to obtain a direct detection of the hot and tenuous gas filling the bubble,
as this is expected to radiate inefficiently (e.g. \citealt{fg12}, \citealt{costa14}; see also next section). 
A promising technique is instead searching for signatures of the inverse Compton effect produced by the hot
electrons in the bubble on the cosmic microwave background (CMB) photons, i.e. the thermal Sunyaev-Zeldovich (tSZ) effect.

Recent works found evidence for the tSZ effect by stacking far-IR and sub-mm data
around large samples of QSOs (e.g. \citealt{ruan15} , \citealt{crichton16}). In particular,
\citet{crichton16} used data from both ACT and Herschel-SPIRE
to build the average far-IR spectral energy distribution of SDSS QSOs in different redshift bins, from
z=0.5 to z=3.5. They found a 3-4$\sigma$ evidence that the average far-IR SEDs of QSOs deviate from what is
expected from pure dust emission, and that these deviations can be explained through the tSZ effect.

The spectral distortions produced by the tSZ effect are normally parameterized by the Compton y-parameter
\beq
y \equiv \int n_e \sigma_T \frac{k_B T_e}{m_e c^2} dl \; ,
\eeq

where $n_e$ and $T_e$ are the density and temperature of the hot electrons, respectively, and the integral is performed along the line of sight.

Because of the limitations in the angular resolution of ACT (FWHM = 1 arcmin, corresponding to a half-light radius of 250 kpc at the median redshift of their sample, z=1.85), \citet{crichton16} could only measure an integrated Compton parameter over the source solid angle, defined as
\beq
Y(z) \equiv d^2_A(z) \int y d\Omega = \frac{\sigma_T}{m_e c^2}\int P_e dV = \frac{2}{3}\frac{\sigma_T}{m_e c^2}E_e \; ,
\eeq

where $d_A(z)$ is the angular diameter distance, i.e. the ratio between the object's physical transverse size and its angular size; $P_e$ is the electron thermal pressure;
and $E_e$ the electron thermal energy. For protons and electrons in thermal equilibrium the total thermal energy in the ionized gas is then
\beq
E_{th}=(1+\mu_e^{-1})E_e = \frac{3}{2} (1+\mu_e^{-1}) \frac{m_e c^2}{\sigma_T} Y(z) \; ,
\eeq

where $\mu_e$ is the mean particle weight per electron (1.17 for primeval gas). \citet{crichton16} measured an average amount of
thermal energy per source of $E_{th}=6.2 (\pm 1.7) \times 10^{60}$ erg. This is about one dex higher
than the thermal energy of the gas heated by the gravitational collapse of their dark matter halos (having
$M_h\lesssim5 \times 10^{12}\; M_{\odot}$, as derived from clustering measurements; \citealt{shen13,efte15}), and was interpreted as evidence
for thermal energy deposition from QSO outflows in their circumgalactic media, i.e. as evidence for large-scale QSO feedback.

Although we are focusing on higher-z QSOs, we are in the position of checking the amount of thermal energy deposited in
our exponentially growing bubbles of hot gas at $t_9$, i.e. the evolutionary time where we expect to observe the source.

By considering that in energy-driven winds about half of the wind kinetic energy goes into bubble thermal energy (e.g. \citealt{w77})
we can integrate Eq.~\ref{input} deriving
\begin{align}
E_{th}\approx \frac{1}{2}E_w(t) = \frac{1}{2}\int_0^{t_9} L_w(t)dt = \frac{v_{w}^2}{4}\int_0^{t_9} \dot m_w (t)dt \\ \notag
& \hspace{-6cm} = \frac{v_{w}^2}{4}\dot m_{w,0}t_{Sal}[e^{t_9/t_{Sal}}-1] \sim 5 \times 10^{60} {\rm erg} \; .
\label{eth}
\end{align}

This value is remarkably similar to that measured by \citet{crichton16}, hence suggesting that AGN outflows are closer to the  energy-driven 
limit than to the momentum-driven limit. This conclusion is in agreement with the findings of \citet{tombesi15} and \citet{feruglio15} based on the comparison between the energy 
of the inner wind and that of the large-scale molecular outflow measured in two nearby QSOs.

\subsection{Cooling of the hot gas in the bubble}

The above computation was performed under the hypothesis that the shocked wind within the bubble does not cool and the outflow is then in the fully energy-driven limit.
We will explore here under which conditions this assumption is satisfied. Assuming that the QSO wind velocity $v_w$ is much higher than the velocity at which the reverse shock is moving,
the post-shock wind temperature, that is the bubble temperature $T_b$, is
\beq
T_b = \frac{3}{16}\frac{\mu m_p}{k_B}v_w^2 \approx 1.2 \times 10^{10} \left(\frac{\mu}{0.59}\right)\left(\frac{v_w}{0.1c}\right)^2 \; \rm{K} \; .
\eeq

For such a hot gas, the only effective cooling mechanisms are free-free (thermal Bremsstrahlung) emission, inverse Compton-scattering, and, for sufficiently compact systems, pair production. The timescale for free-free cooling can be estimated as \citep{mvw10,costa14}
\beq
t_{ff} \approx 8 \times 10^{5} \left(\frac{T_b}{10^{10}\rm{K}}\right)^{1/2}\left(\frac{n_e}{10^{-3}\rm{cm}^{-3}}\right)^{-1} \; \rm{Myr} \; ,
\eeq

which is obviously much longer than the Hubble time for the temperatures and densities ($n_e\sim10^{-5}-10^{-7}$ cm$^{-3}$) of our systems. 

As the hot gas in the bubble is exposed to the radiation field of the QSO, the high-energy electrons in the bubble can exchange energy with the QSO photons through Compton scattering. If the temperature of the system is higher/lower than the QSO Compton temperature ($T_C\sim2\times10^4$~K, e.g. \citealt{sazonov05}), which depends only on the QSO spectral shape, the gas can be cooled/heated. The relevant timescale for Compton cooling/heating can be written as
\beq
t_C \approx 10^2 \left(\frac{R}{1\rm{kpc}}\right)^2 \left(\frac{M_{BH}}{10^8 M_{\odot}}\right)^{-1}\lambda^{-1} \frac{T_b}{|T_b-T_C|}\; \rm{Myr} \; .
\eeq

For our system, $T_b >> T_C$ and the bubble might cool because of inverse Compton scattering. However, when putting in the above 
equation the values of the bubble radius $R(t)$, and BH mass $M_{BH}(t)$ at any given accretion time, the timescales for Compton cooling are from 30 to 1000 times larger than the flow time 
for the whole halo mass range explored in the previous sections. We therefore conclude that Compton scattering of the QSO photons is an inefficient mechanism to cool down the bubble.

At high redshift, however, the CMB provides a large reservoir of photons 
available for Compton scattering because its energy density scales as $(1+z)^4$. Since the CMB temperature $T_{CMB}\sim 2.73(1+z)\;\rm{K}<<T_b$, 
the hot electrons in the bubble might cool efficiently by inverse Compton scattering on CMB photons (IC-CMB). Following \citet{mvw10}, for the cosmology adopted here and at $z\gtrsim2$, the cooling time for the IC-CMB process can be approximated with
\beq
t_{IC-CMB}/t_U\sim 134 (1+z)^{-5/2} \; ,
\eeq

where $t_U(z)$ is the  age of the Universe at redshift $z$. By considering that $t_U(z) = t_{acc} + t_U(12.5)$ and inserting it in the above equation, we find that 
$t_{IC-CMB}$ is always larger than $t_{acc}$, but they are of the same order for $t_{acc}\gtrsim100$ Myr. This means that at high-z the IC-CMB is an effective cooling mechanism
for the hot gas in the bubble. This may produce significant departures from the pure energy-driven limit discussed above, but we defer  a detailed treatment of the outflow  behaviour to future hydrodynamic
simulations which include IC-CMB cooling.

We finally discuss the possibility that the high-temperature plasma within the bubble ($T_b\sim10^{10}$~K) may cool down because
of pair production. Indeed, if the system is sufficiently compact, high-energy photons may produce electron-positron pairs that can slow
down the fast protons in the plasma through Coulomb interactions, effectively cooling down the bubble \citep{begelman87}. 
This would happen if the system is optically thick to pair production, 
i.e. when the dimensionless compactness parameter $l\equiv L_b \sigma_T / R m_e c^3>>1$ \citep{lz87}. 
Here $L_b$ is the free-free luminosity (see e.g. Eq. B1 in \citealt{fg12}) at $\sim$MeV energies of the thermal plasma in the bubble and 
$R$ is the bubble radius. For the values typical  of our systems, $l\sim10^{-13}$ at any given accretion time, making pair production an inefficient cooling mechanism.

\subsection{Stability analysis and shell fragmentation}

We provide here considerations on the stability and fragmentation of the gas shell pushed by the QSO. We consider the case of a BH starting from a seed mass of $10^4\;M_{\odot}$
at $z=12.5$ placed within a dark matter halo growing to $M_h=3\times10^{11}\;M_{\odot}$ at $z=6$, for which the outflow is always supersonic (Fig.~\ref{quad} right).

As opposed to the case treated in Sections 4 and 5 of a QSO bubble expanding in the IGM where the effects of gravity are negligible, for an outflow expanding within a dark matter halo
gravity combines with the acceleration of the expanding shell in generating Rayleigh-Taylor (RT) instabilities. In particular, for a fluid with a radial acceleration $\ddot R$, 
the growth rate of the perturbations at a given spatial scale $\lambda$ is $\Gamma_{RT}=\sqrt{2\pi g' /\lambda}$, where $g'=\ddot R + g$ is the apparent gravitational, or net radial, acceleration of the system \citep{drazin02}. The growth rate of the perturbations is  faster at the smallest scales, i.e. those comparable with the shell 
thickness $\Delta R\sim R/(3+3\mathcal{M}^2)$, where $R$ is the bubble radius and $\mathcal{M}$ is the outflow Mach number \citep{w77}.

For the case discussed here, $g'>0$ for $t_{acc}> 5$ Myr, and the e-folding time of the small-scale perturbation growth $t_{RT}\equiv1/\Gamma_{RT}$ becomes shorter than $t_{acc}$ soon afterwards. Therefore, RT instabilities have enough time to develop and alter the whole structure of the expanding shell.
The detailed structure and fragmentation history of the shell can be studied only with high-resolution simulations. We  assume here that the overall spherical structure of the
shell is preserved even at late accretion times, and hence its expansion history presented in the previous section holds. The hydrodynamics
simulations performed by \citet{costa14} assuming a constant source of energy and mass show that, after developing RT instabilities, the global structure of an expanding shell in energy-driven QSO outflows is preserved
even at late times and can be described with good accuracy by simple numerical and analytic methods.

\begin{figure}[t]
\includegraphics[angle=0, width=8.7cm]{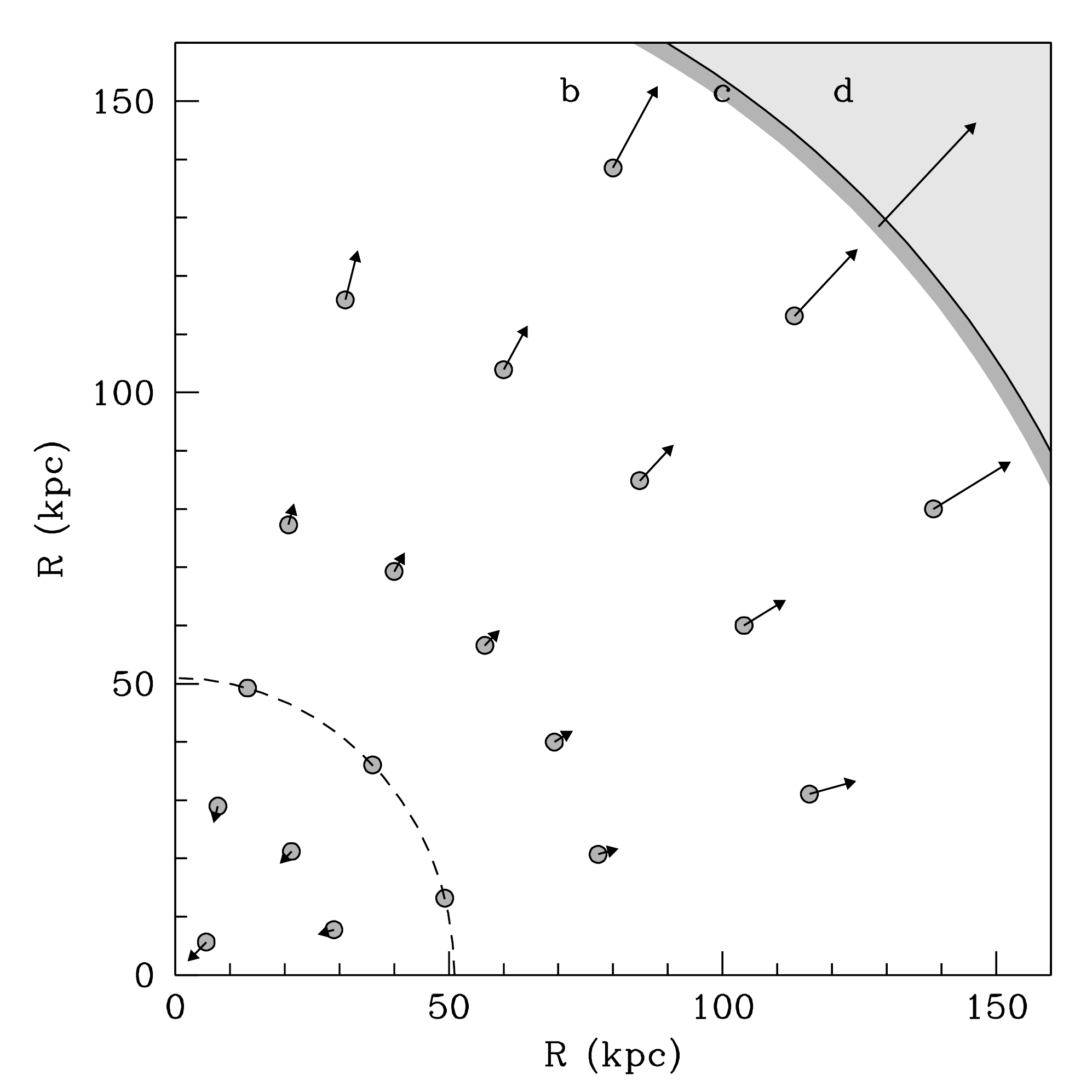}
\caption{Schematic view (to scale) of the outflow structure around a QSO at z=6. Regions $b, c$, and $d$ are as in Fig.~\ref{ofs}. The host halo mass was fixed to $3\times 10^{11}\;M_{\odot}$. As Rayleigh-Taylor instabilities develop, 
the gas shell loses fragments (clouds) of the size of its thickness. Depending on when a cloud detaches from the shell, it will be seen as an outflowing or inflowing cloud, 
where ``younger'' clouds have larger distances from the QSO and higher receding velocities. 
The length of each arrow is proportional to the cloud velocity. The dashed curve at $\sim 50$ kpc shows the boundary where clouds are seen as infalling or outflowing. Clouds that have detached from the shell at $t_{acc}<186$ Myr have already fallen back to the BH by the time the system is 
observed ($t_9$) and may accumulate and produce large reservoirs of cold gas around the QSO.} 
\label{map}
\end{figure}

At any given time, the first fragments (clouds) separating from the expanding shell will be those whose sizes are similar to the shell thickness $\Delta R$.
Once detached from the outflowing shell, the clouds are subject to the gravitational field of the halo. Because
of the high contrast between the gas density within the cloud $\rho_{cl}$ and that of the hot tenuous gas within the bubble $\rho_b$  ( $\chi\equiv\rho_{cl}/\rho_b \sim 10^{4-5}$), the ram
pressure exerted by the hot gas is not sufficient to slow down the cloud, which eventually falls back towards the BH in a free fall-time
$\tau_{ff}=(v_s+\sqrt{v_s^2+2gR})/g$, where $v_s$, $R$, and $g$ are the shell velocity, the bubble radius and the acceleration of gravity on the shell, respectively, at the time of the cloud detachment $t_{det}$. If the free-fall time $\tau_{ff}$ is shorter than the cloud age $\tau\equiv t_9-t_{det}$, by the time we observe the system the cloud has already fallen back 
to the BH, otherwise it may be observed. For the system considered here, this means that only clouds detaching from the shell at $t_{det}>186$ Myr can be observed. At these late times, the shell thickness varies only between 2 and 3 kpc, so 
clouds with radii $r_{cl}=\Delta R /2 \sim 1.0-1.5$ kpc are expected. Given the gas density within the cloud $\rho_{cl}=\rho_{shell}=(1+\mathcal{M}^2)\rho_{gas}\sim 10^{-26}$~g cm$^{-3}$, this corresponds 
to gas masses of $M_{cl}\sim 1-2 \times10^6 \; M_{\odot}$ per cloud. Also, each cloud detaching at $t_{det}>186$ Myr should have the following velocity and radius
by $t_9$: $v_{cl}(t_9)=v_s(t_{det}) -g\tau$ and $R_{cl}=R_b +v_s\tau-g\tau^2/2$. Clouds that detach from the shell at late times, should be seen close to the shell itself and with 
large outflowing velocities (see Fig.~\ref{map}). Clouds that detach from the shell at times just after 186 Myr, should be seen close to the BH and with infalling velocities. 
For the case considered here, the transition between the regions populated by outflowing and infalling clouds occurs at $R_{cl}=51$ kpc, where clouds at zero velocities should be observed
(see Fig.~\ref{map}).

Individual clouds are expected to suffer from radiative losses and cool down  following the cooling function \citep{cioffi91} 
\beq
\Lambda = \left\{
\begin{array}{lr}
2.49 \times 10^{-27} T_{cl}^{1/2} {\rm ~erg~cm^3~ s^{-1}}& (free-free)\\
1.3 \times 10^{-19}\zeta T_{cl}^{-1/2} {\rm ~erg~cm^3~ s^{-1}}& (metals)\\
\end{array}
\right. ,
\eeq

where the first relation is valid for gas made only of hydrogen and helium and the second is valid for a gas with $10^5<T<10^7$~K and metallicity $\zeta$ relative to solar.
The corresponding cooling times $\tau_{cool}\sim 1.5k_BT_{cl}/(n_{cl}\Lambda)$ are then
\beq
\tau_{cool} \sim \left\{
\begin{array}{lr}
2.6  \left(\frac{T_{cl}}{10^6 K}\right)^{1/2} n_{cl}^{-1} {\rm ~Myr}& (free-free)\\
0.05 \left(\frac{T_{cl}}{10^6 K}\right)^{3/2} (n_{cl}\zeta)^{-1}  {\rm ~Myr}& (metals)\\
\end{array}
\right. .
\eeq

For the system considered here, where $T_{cl}=T_{gas}\sim 0.3-1.0 \times 10^{6}$~K (assuming $T_{gas}=T_{vir}$) and $n_{cl}\sim 0.01$ cm$^{-3}$, it follows that
$t_{cool}<<t_{acc}$ only if $\zeta\gtrsim 0.05$, i.e. the clouds would cool significantly by the time we can observe them only if a significant fraction of metals is already
present in the ambient gas. We speculate here that those clouds that detach early from the shell and fall back towards the black hole may both 
provide fuel for further BH accretion, in analogy with chaotic cold accretion models \citep{gtb17}, and a reservoir of ``cold'', $T\sim 10^4$~K gas, like the one probed by MUSE through the ubiquitous detection of giant Lyman $\alpha$ halos extending for a few tens of kpc around luminous QSOs at $3<z<4$
\citep{borisova16}. 

As the clouds move within the hot gas in the bubble they are subject to Kelvin-Helmoltz (KH) instabilities, which can effectively remove material from the surface of the clouds until their disruption (see also \citealt{fs16}).
The mass loss is particularly relevant at scales $\lambda \sim r_{cl}$, for which the growth timescale of the KH instability is \citep{murray93}
\beq
\tau_{KH}=\lambda \frac{\rho_{cl} + \rho_b}{(\rho_{cl} \rho_b)^{1/2}v_{cl}} \approx \frac{\chi^{1/2} r_{cl}}{v_{cl}}
\eeq

in the limit of a high-density contrast between the density of the cloud and that of the bubble, $\chi>>1$, as applicable here.

Since the cloud loses mass at a rate $\dot M_{cl} \sim 4\pi \rho_{cl} r_{cl}^3/\tau_{KH}$, the characteristic stripping time for the cloud mass is
\beq
\tau_{strip} = \frac{M_{cl}}{\dot M_{cl}} = \frac{\tau_{KH}}{3} ,
\eeq

where $M_{cl}=4\pi \rho_{cl} r_{cl}^3/3$ is the mass of the cloud \citep{lm00}.

For the observable clouds in the system discussed above, the stripping time $\tau_{strip}$ is of the same order of the cloud age at the time of the observation. Therefore, individual clouds
may survive the KH instabilities and be observed. This conclusion is reinforced by considering that the mass of those clouds that detach from the shell soon after $t_{det}$, and hence that move 
at relatively low velocities ($\sim 100$ km s$^{-1}$), is close to the self-gravitating critical mass required to remain stable against KH instabilities
$M_{cr}\sim (6^{1/2}\pi v_{cl}^3)/(G^{3/2}\chi^2 \rho_b^{1/2})\sim 7\times 10^6\; M_{\odot}$ \citep{murray93}.
Because of the high temperature in the bubble, the clouds are subject to significant evaporation \citep{cowie77,marcolini05}.
The mass evaporation rate can be written as
\beq
\dot M_{ev} = \left\{
\begin{array}{lr}
1.36 \dot M_{cl} \sigma_0^{-5/8}  & \rm{if} \; \sigma_0 \geq 1\\
\dot M_{cl} & \rm {if} \; \sigma_0 < 1\\
\end{array}
\right. , 
\eeq

where the dimensionless parameter $\sigma_0$ is defined as
\beq
\sigma_0 = 4.22 \times 10^{-3} \left(\frac{T_b}{10^6 \rm{K}}\right)^2 n_b^{-1} \left(\frac{r_{cl}}{1 \rm{pc}}\right)^{-1} \; ,
\eeq

and 
\beq
\dot M_{cl}= 4.34 \times 10^{-7} \left(\frac{T_b}{10^6 K}\right)^{5/2} \left(\frac{r_{cl}}{1 \rm{pc}}\right)\; M_{\odot} \rm{yr}^{-1} \; .
\eeq

In our case, $\sigma_0>>1$ and the evaporation time can be written as
\beq
\tau_{ev} = \frac{M_{cl}}{\dot M_{ev}}\sim0.03\left(\frac{T_b}{10^{10}K}\right)^{-5/4} \left(\frac{r_{cl}}{1 kpc}\right)^{11/8} \left(\frac{\chi}{5000}\right)^{5/8} \left(\frac{n_{cl}}{0.01}\right)^{3/8}Myr \; .
\eeq

Therefore, all clouds should quickly evaporate if $T_b\sim 10^{10}$~K. However, as discussed in the previous section, for bubbles produced by high-z QSOs, IC-CMB cooling would
be effective in bringing the plasma to lower temperatures. For $T_b\sim$ a few $\times 10^7K$, the evaporation time would become as long as the cloud age at $t_9$. Therefore, an intriguing prediction of this simple model is that a distribution of clouds around QSOs like the one shown in Fig.~\ref{map} can be only observed in systems at $z\sim 6$ and beyond.
We defer  a detailed treatment of the shell fragmentation and formation of gas clouds to future high-resolution hydrodynamical simulations.

\section{Conclusions}

We investigated the physics and time evolution of large-scale outflows produced by early QSOs powered by exponentially growing black holes.
We assumed that these systems grow to $M_{BH}=10^9\;M_{\odot}$ by $z=6$ by accreting at the Eddington limit and converting a fixed fraction of their bolometric output into a wind. 
This means that the outflow source power is also growing exponentially.
We first considered the cases of energy- and momentum-driven outflows expanding in a region where the gas and total mass densities are uniform and equal to 
the average values in the Universe at $z\geq6$. We then extended our computations to the case of QSOs placed at the centre of early dark matter halos
of different masses and starting from different seed BH masses. We made considerations on the energetics of the outflow, on the cooling of the hot gas in the QSO-inflated bubble, 
and on the stability and structure of the expanding gas shell. Our main results can be summarized as follows:

$\bullet$
For a SMBH/QSO growing in mass/power with an e-folding (Salpeter) time $t_{Sal}$, the late time expansion of the bubble radius
 is also exponential, with an e-folding time of $5t_{Sal}$ and $4t_{Sal}$ for an energy-driven and a momentum-driven outflow,  respectively. 
In the case of a QSO expanding within a uniform density field we provided analytic solutions to the time evolution of the bubble radius. 

$\bullet$
For a QSO outflow expanding within a field where the gas and total mass densities are uniform and equal to the average field values at $z>6$, the expansion of
the bubble is only affected by the gas density, whereas the gravitational drag exerted by dark matter is negligible. The latter is instead relevant for 
outflows produced by QSOs at the centre of large dark matter halos. 

$\bullet$ 
We considered energy-driven outflows produced by black holes growing from seeds with a mass range of $10^3-10^6M_{\odot}$ and placed within growing dark matter halos 
spanning a mass range of $3\times10^{11}-10^{13}\;M_{\odot}$ at $z=6$. We followed the evolution of the source power and of the gas and dark matter density profile in the halos from the
beginning of the accretion until $z=6$. For a given final BH mass ($10^9\;M_{\odot}$ in our case), the bubble radius and velocity at z=6 do not depend on the initial seed mass: a bubble inflated by a smaller 
(larger) seed simply  takes more (less) time to grow to the same final value $R(t_9)$. The final bubble radius and velocity are instead smaller for larger halo masses. At z=6, bubble
radii in the range 50-180 kpc and velocities in the range 400-1000 km~s$^{-1}$ are expected for QSOs hosted by halos in the mass range $3\times10^{11}-10^{13}\;M_{\odot}$.
These radius and velocity scales compare well with those measured for the outflowing gas in the well-known z=6.4 QSO SDSS~J1148+5251.

$\bullet$
We assumed that the gas in the halo is at the virial temperature. For large enough halos, where the gas temperature and sound speed are higher,  the expansion of the 
bubble may become subsonic in a given time interval. We find that for halos with $M_h\leq3\times10^{11}\;M_{\odot}$ at z=6, the outflow is always supersonic. The fraction of time spent at subsonic velocities increases for larger masses, until it is always subsonic for $M_h\geq5\times10^{12}\;M_{\odot}$ at z=6. We also explored the effects of  
assuming a lower ambient gas temperature, down to $T_{vir}/4$. 
We found that the halo mass thresholds for fully subsonic and fully supersonic outflows do not strongly depend on the assumed gas temperature and corresponding density profile. 
For lower temperatures and steeper profiles, the bubble radii and velocities can reach values up to 1 Mpc and a few $\times 1000$ km~s$^{-1}$, respectively.

$\bullet$
In the case of an energy-driven outflow, we computed a total thermal energy of $\sim5 \times 10^{60}$ erg contained in the bubble around the QSO.
This number is in excellent agreement with the value of $(6.2\pm 1.7) \times 10^{60}$ erg per QSO as derived from a large sample of QSOs through the detection of the thermal Sunyaev-Zeldovich effect in their stacked far-IR spectra. This suggests that QSO outflows are closer to the energy-driven limit than to the momentum-driven limit.

$\bullet$
We investigated the stability of the expanding gas shell in the case of an energy-driven supersonic outflow propagating within a dark matter halo. We found that the shell is Rayleigh-Taylor
unstable already at $t>5$ Myr and, by means of a simple model, we investigated the fate of the fragments detaching from the shell. We found that these fragments should rapidly evaporate because of the extremely high temperature of the hot gas bubble, unless this cools effectively. Since the only effective cooling mechanism for such a gas is inverse Compton by the CMB photons (IC-CMB), which is important only at $z\geq 6$, we speculate that such shell fragments can be observed only around high-z QSOs, where IC-CMB cooling of the bubble gas can prevent their evaporation.

$\bullet$
We finally propose that those shell fragments that have already fallen back towards the centre of the dark matter halo by the time we
observe the QSO may accumulate and constitute a reservoir
of relatively cold gas ($T\sim 10^4$ K) on scales of up to a few tens of kpc. This mechanism could explain the ubiquitous presence of such a gas observed by MUSE
around $z\sim3.5$ QSOs.

\begin{acknowledgements}
We acknowledge stimulating discussions with M. Gaspari, L. Ciotti, M. Meneghetti, S. Ettori, and A. Negri, and a clear and constructive report from the referee.
We acknowledge financial contribution from the agreement ASI-INAF I/037/12/0.
\end{acknowledgements}

\bibliographystyle{aa} % style aa.bst

\bibliography{/Users/gilli/bubbles/biblio}

\end{document}